\documentclass[12pt]{article}
\usepackage{amsmath, amssymb, booktabs, longtable, geometry}
\usepackage{algorithmic}
\usepackage{appendix}
\usepackage{listings}
\usepackage{xcolor}
\usepackage{multirow}
\usepackage{array}
\usepackage{graphicx}
\usepackage{float}
\usepackage{svg}
\usepackage{bm}
\usepackage[authoryear,round,sort&compress]{natbib}
\usepackage[hidelinks]{hyperref}
\providecommand{\figalttext}[2][]{}
\graphicspath{{Fig/}}

\title{Identifying Damage Pathways Linking Sequence Composition to Storage Failure in DNA Data Storage via High-Dimensional Mediation Analysis}

\author{
	\small
	Jingyi Li$^{1}$, Huaming Wu$^{2}$,
	Haixiang Zhang$^{1,}$%
	\thanks{Corresponding author. E-mail: haixiang.zhang@tju.edu.cn}\\[-0.1cm]
	\footnotesize
	$^{1}$School of Mathematics and KL-AAGDM, Tianjin University,
	Tianjin 300350, China\\[-0.05cm]
	\footnotesize
	$^{2}$Center for Applied Mathematics, Tianjin University,
	Tianjin 300072, China
}

\date{}

\begin{document}
	
	\maketitle
	
	\begin{abstract}
		\textbf{Motivation:} DNA data storage offers extraordinary information density and long-term durability, but its reliability is limited by sequence-dependent errors introduced during synthesis and accumulated during storage. It remains unclear how sequence composition is associated with storage failure through specific molecular damage components. 
		
		\textbf{Results:} We develop a high-dimensional semiparametric mediation framework for survival outcomes. GC content is treated as the exposure, a high-dimensional baseline damage spectrum (a vector of per-read damage counts stratified by trinucleotide context and error type) as the mediator, and storage-quality failure as the outcome. Nonlinear covariate effects in both the mediator and survival models are approximated using deep neural networks. A three-step procedure combining product-of-coefficients screening, Smoothly Clipped Absolute Deviation (SCAD) penalized estimation, and joint significance testing is developed for mediator selection and inference. Applied to an aging experiment on electrochemically synthesized DNA, the method identifies 14 significant mediators, all corresponding to single-base deletions, with estimated mediated effects concentrated in trinucleotide contexts ending in C. These results reveal deletion-type damage as a major pathway linking sequence composition to reduced archival reliability and suggest candidate sequence features for future optimization and error-control strategies. The proposed framework thus offers a mechanism-oriented statistical approach for understanding and improving the reliability of DNA data storage.
		
		\textbf{Availability:} The software is publicly available at: \url{https://github.com/cvvvfvfvf/deep-hima}.
		
	\end{abstract}
	
	\noindent\textbf{Keywords:} DNA storage, Cox model, High-dimensional mediation analysis, Deep learning
	
	\section{Introduction}
	Analysts estimate that global memory demand will reach approximately $3\times10^{24}$ bits by 2040 \citep{zhirnov2016}. Conventional storage media, such as magnetic hard drives, suffer from limited density, high energy consumption for maintenance, and a physical lifespan measured in decades. At the theoretical maximum, single-stranded DNA can encode up to 455 exabytes per gram, and DNA-encoded information can remain readable despite degradation over millennia \citep{church2012, goldman2013}. Early studies demonstrated the feasibility of encoding and recovering megabit-scale digital information in oligonucleotide pools, including a 5.27-megabit book \citep{church2012, goldman2013}, and subsequent work scaled this to hundreds of megabytes with random-access retrieval \citep{organick2018}.
	
	Despite this promise, the widespread adoption of DNA storage remains hindered by two interrelated challenges. The synthesis process itself introduces a substantial baseline error burden that depends strongly on sequence composition \citep{chen2020quantifying,gimpel2023}, arising primarily from coupling failures, which produce deletions, and from depurination, which leads to substitutions \citep{yeom2023}. In addition, damage that progressively degrades sequence readability accumulates during aging under ambient conditions \citep{lindahl1993}. Error-correcting codes (ECC) have been the dominant engineering response to these challenges. HEDGES~\citep{press2020}, for instance, demonstrates that error-free recovery is achievable only up to a limited DNA error rate, beyond which decoding fails.
	
	However, existing reliability analyses in DNA storage have primarily focused on coverage-based metrics. Studies have characterized how extreme guanine and cytosine (GC) content leads to under-representation in sequencing libraries through polymerase chain reaction (PCR) amplification bias \citep{aird2011}, and how sequence constraints can reduce coverage dispersion in synthesis pools \citep{gimpel2023}. These investigations address whether a sequence is present at sufficient copy number for retrieval (the dropout problem), but do not directly quantify how the baseline damage spectrum of individual molecules affects their retrievability. Our objective is therefore to identify which specific trinucleotide contexts and error types of baseline molecular damage are associated with accelerated loss of oligo recoverability during aging.
	
	To address this question, we adopt a mediation analysis framework \citep{mackinnon2002}, combining a Cox proportional hazards model for storage-quality failure \citep{cox1972} with a high-dimensional mediator. GC content is the exposure, known to affect synthesis efficiency and error profiles \citep{yeom2023,gimpel2023}. We define the baseline damage spectrum as a vector of per-read damage counts stratified by trinucleotide context and error type, including transition (Ti), transversion (Tv), deletion (del), insertion (ins), and deletion run (delrun), measured after synthesis (day 0). By decomposing the total effect of GC content into a direct path and an indirect path mediated by this spectrum, we aim to identify which specific components are most consequential for archival reliability. Practically, if certain error-prone contexts dominate the mediated pathway, the decomposition provides concrete targets for sequence design and ECC parameterization.
	
	High-dimensional mediation analysis has emerged as a powerful tool for disentangling complex causal pathways in omics studies. In epigenetic research, \citet{zhang2016hdm} developed methods for estimating and testing mediation effects when the number of mediators far exceeds the sample size, using penalized regression and multiple testing procedures to control the false discovery rate (FDR). Subsequent work has extended these methods to handle quantile-based effects \citep{zhang2024quantile}, composite null hypotheses under correlated mediators \citep{huang2019composite}, and Bayesian frameworks that jointly estimate sparse mediation structures \citep{song2020bayesian}. These advances have enabled researchers to identify specific molecular pathways through which environmental exposures affect disease outcomes, even when the mediator space is vast and only a small fraction of mediators are truly active. Recent work has addressed high-dimensional mediation with survival outcomes, applying a least absolute shrinkage and selection operator (LASSO) screening followed by joint significance testing \citep{zhang2021}, and Bayesian shrinkage for sparse effect estimation \citep{song2020bayesian}. These approaches, however, assume either a linear mediator model or a parametric nonlinear form for the confounding adjustment, which may be misspecified when covariate effects are complex. In the DNA storage context, the relationship between sequence composition covariates (e.g., homopolymer load, trinucleotide entropy, GC heterogeneity) and the damage spectrum is unlikely to be well captured by a linear model.
	When the outcome of interest is a time-to-event variable, the mediation framework must be adapted to handle censoring and the semiparametric structure of survival models.
	
	In this paper, we propose a semiparametric mediation framework for investigating how GC content is associated with DNA storage reliability through a high-dimensional spectrum of molecular damage. The proposed framework accommodates a survival outcome, a large number of candidate mediators, and potentially nonlinear effects of covariates, with the latter approximated by DNNs. To identify mediating damage components, we develop a three-step procedure consisting of product-of-coefficients screening, SCAD-penalized alternating estimation, and joint significance testing. The main contributions are threefold. First, we formulate DNA storage degradation as a high-dimensional mediation problem with a survival outcome, linking sequence composition, baseline molecular damage, and loss of recoverability beyond conventional coverage-based analyses. Second, we develop a semiparametric procedure that combines flexible nonlinear covariate adjustment with sparse selection and inference for numerous context- and error-specific mediators. Third, the framework provides mechanism-oriented results by identifying the error types and trinucleotide contexts through which sequence composition is associated with reduced archival reliability, rather than merely predicting oligo failure.
	
	The remainder of the paper is organized as follows. Section~\ref{sec:framework} presents the proposed semiparametric mediation framework, together with the estimation and inference procedure. Section~\ref{sec:simulations} investigates the finite-sample performance of the proposed method through simulation studies. Section~\ref{sec:application} applies the method to DNA storage data and presents the resulting mediation analysis. Section~\ref{sec:conclusion} concludes the paper. Additional simulation results are reported in the Supplementary Material.
	
	\section{Semiparametric Mediation Framework for DNA Storage Reliability}\label{sec:framework}
	\subsection{Model formulation and mediation pathways}
	DNA data storage reliability is determined by a sequence of interconnected
	processes rather than by a single source of error. The designed nucleotide
	sequence first influences the fidelity of DNA synthesis and the resulting
	baseline error profile, while these synthesis-associated defects may in turn
	affect the ability of an oligonucleotide to remain reliably recoverable during
	subsequent storage. Therefore, an observed association between sequence
	composition and storage failure may arise through at least two routes: a direct
	association that is not explained by the measured baseline damage profile, and
	an indirect pathway operating through specific types of sequence-dependent
	damage. Distinguishing these pathways is important because it moves the analysis
	beyond identifying sequence features associated with poor storage performance
	and toward determining which molecular damage components may carry that
	association.
	
	These considerations motivate a general mediation framework for studying how
	sequence composition is associated with DNA storage reliability through
	intermediate damage processes. For the $i$th oligonucleotide, let
	$X_i\in\mathbb{R}$ denote a scalar sequence-composition feature of primary
	interest, $Z_i=(Z_{i1},\ldots,Z_{iq})^\top\in\mathbb{R}^q$ denote additional
	sequence-derived characteristics, and
	$M_i=(M_{i1},\ldots,M_{ip})^\top\in\mathbb{R}^p$ denote a potentially
	high-dimensional baseline damage profile. 
	
	In our DNA storage application, $X_i$ is taken to be the GC content of the
	design sequence $s_i$, while $Z_i$ includes additional sequence characteristics
	such as the maximum homopolymer length, homopolymer load, 3-mer Shannon entropy,
	and GC heterogeneity. The mediator vector $M_i$ represents the baseline damage
	spectrum measured immediately after synthesis. Each component is indexed jointly
	by a local sequence context and an error type. Specifically, we consider the
	64 trinucleotide contexts together with five error categories (Ti, Tv, del, ins, and delrun), giving $p=64\times 5=320$ candidate mediators. 
	
	Let $T_i$ denote the storage-quality failure time for oligonucleotide $i$. We observe
	\[
	Y_i=\min(T_i,C_i), \qquad
	\Delta_i=I(T_i\leq C_i),
	\]
	where $\Delta_i=1$ indicates an observed storage-quality failure and
	$\Delta_i=0$ indicates censoring. Given discrete follow-up times, the event time is recorded as the first assessment at which the failure criterion is met. The specific observation
	schedule and failure criterion used in our DNA aging experiment are described
	in Section~\ref{sec:application}.
	
	To characterize the mediation mechanism, we propose a high-dimensional partially linear mediation model with a survival outcome. The model jointly links sequence composition, the baseline damage spectrum, and storage-quality failure through two interconnected components. The first component describes how the exposure and individual damage components are associated with the storage-failure hazard, whereas the second characterizes how the exposure affects each component of the high-dimensional damage spectrum after adjustment for other sequence characteristics. Specifically, we propose:
	
	\begin{equation}
		\lambda\bigl(t|X,\mathbf{M},\mathbf{Z}\bigr)
		= \lambda_{0}(t)\exp \{
		\gamma X + \boldsymbol{\beta}'\mathbf{M} + g(\mathbf{Z})
		\},
		\label{eq:survival-model}
	\end{equation}
	\begin{equation}
		M_k = \alpha_{k}X+g_{k}(\mathbf{Z})+e_{k}.
		\label{eq:mediator-model}
	\end{equation}
	
	Here, $\lambda(t\mid X,\mathbf{M},\mathbf{Z})$ denotes the conditional hazard function of $T$ given $X$, $\mathbf{M}$, and $\mathbf{Z}$. $X$ represents sequence composition, taken as GC content in our application, and $\mathbf{M}=(M_1,\ldots,M_p)^\top$ represents the baseline damage spectrum. In Equation~\eqref{eq:survival-model}, $\gamma$ captures the effect of GC content on storage failure that is not explained by the measured damage spectrum, whereas $\beta_k$ measures the association between the $k$th damage component and failure risk. In Equation~\eqref{eq:mediator-model}, $\alpha_k$ quantifies how GC content affects the $k$th damage component. The product \(\alpha_k\beta_k\) quantifies the strength and direction of the \(X\to M_k\to T\) pathway on the log-hazard scale. The unknown functions $g(\mathbf{Z})$ and $g_k(\mathbf{Z})$ flexibly adjust for potentially nonlinear effects of other sequence characteristics. The term $e_k$ denotes the statistical residual for the $k$th damage component, representing oligo-to-oligo variation in $M_k$ that remains after accounting for GC content $X$ and the other sequence characteristics $\mathbf{Z}$. Let $\mathbf{e}=(e_1,\ldots,e_p)^\top$, where \(e_k\) is a zero-mean random error term. The correlation among the damage components is accommodated through $\operatorname{Cov}(\mathbf{e})=\Sigma_e$. 
	
	\subsection{Counterfactual interpretation of the mediation effect}
	\label{subsec:causal_interpretation}
	
	We next provide a counterfactual interpretation of the pathway coefficient
	$\alpha_k\beta_k$. The purpose of this subsection is to clarify the scale on
	which the product-of-coefficients quantity admits a causal interpretation. For an exposure level $x$, let $M_k(x)$
	denote the potential value of the $k$th mediator that would be observed if
	the exposure were set to $x$. Let $
	T(x,\bm m)$
	denote the potential failure time that would be observed if the exposure
	were set to $x$ and the mediator vector were externally set to
	$\bm m=(m_1,\ldots,m_p)^\top$.
	
	Under the mediator model $M_k=\alpha_k X+g_k(\bm Z)+e_k,$ the potential value of the \(k\)th mediator is given by $M_k(x)=\alpha_k x+g_k(\bm Z)+e_k.$
	Here, $\mathbf Z$ is held fixed, and $e_k$ represents the same exogenous
	disturbance across different counterfactual exposure levels, as implied by
	the structural equation representation. Therefore, for any two exposure
	levels $x$ and $x^*$
	\begin{equation}
		M_k(x)-M_k(x^\ast)
		=
		\alpha_k(x-x^\ast).
		\label{eq:mediator_contrast}
	\end{equation}
	
	For the survival outcome, suppose that the counterfactual hazard under an
	intervention $(X,\bm M)=(x,\bm m)$ follows
	\begin{equation}
		\lambda\{t\mid x,\bm m,\bm Z\}
		=
		\lambda_0(t)
		\exp
		\left\{
		\gamma x
		+
		\sum_{j=1}^p\beta_jm_j
		+
		g(\bm Z)
		\right\}.
		\label{eq:counterfactual_cox}
	\end{equation}
	The model contains no exposure--mediator interaction. Therefore, the effect
	of changing the $k$th mediator is multiplicative on the hazard scale and
	additive on the log-hazard scale.
	
	To isolate the pathway through mediator $M_k$, define the hybrid
	counterfactual mediator vector
	\begin{multline*}
		\bm M^{(k)}(x,x^\ast)
		=
		\bigl\{
		M_1(x^\ast),\ldots,M_{k-1}(x^\ast),M_k(x),\\
		M_{k+1}(x^\ast),\ldots,M_p(x^\ast)
		\bigr\}^\top.
	\end{multline*}
	Thus, only the $k$th mediator is allowed to change from its natural value
	under $x^\ast$ to its natural value under $x$, whereas all the remaining
	mediators are kept at their natural values under $x^\ast$.
	
	We define the path-specific natural indirect effect through $M_k$ on the
	log-hazard scale by
	\begin{equation*}
		\operatorname{NIE}^{\log\lambda}_k
		(t;x,x^\ast)
		=
		\log
		\frac{
			\lambda
			\left[
			t\mid
			x,\bm M^{(k)}(x,x^\ast),\bm Z
			\right]
		}{
			\lambda
			\left[
			t\mid
			x,\bm M(x^\ast),\bm Z
			\right]
		}.
		\label{eq:nie_loghazard_def}
	\end{equation*}
	Using the proportional hazards model in
	\eqref{eq:counterfactual_cox}, the numerator is
	\begin{align}
		&
		\lambda
		\left[
		t\mid
		x,\bm M^{(k)}(x,x^\ast),\bm Z
		\right]
		\nonumber\\
		&\quad=
		\lambda_0(t)
		\exp
		\Bigg\{
		\gamma x
		+
		\beta_k M_k(x)
		+
		\sum_{j\neq k}\beta_jM_j(x^\ast)
		+
		g(\bm Z)
		\Bigg\},
		\label{eq:nie_num}
	\end{align}
	and the denominator is
	\begin{align}
		&
		\lambda
		\left[
		t\mid
		x,\bm M(x^\ast),\bm Z
		\right]
		\nonumber\\
		&\quad=
		\lambda_0(t)
		\exp
		\Bigg\{
		\gamma x
		+
		\beta_k M_k(x^\ast)
		+
		\sum_{j\neq k}\beta_jM_j(x^\ast)
		+
		g(\bm Z)
		\Bigg\}.
		\label{eq:nie_den}
	\end{align}
	Taking the ratio of \eqref{eq:nie_num} and \eqref{eq:nie_den}, all terms
	other than the $k$th mediator cancel, giving
	\begin{align*}
		\frac{
			\lambda
			\left[
			t\mid
			x,\bm M^{(k)}(x,x^\ast),\bm Z
			\right]
		}{
			\lambda
			\left[
			t\mid
			x,\bm M(x^\ast),\bm Z
			\right]
		}
		&=
		\exp
		\left[
		\beta_k
		\left\{
		M_k(x)-M_k(x^\ast)
		\right\}
		\right]
		\nonumber\\
		&=
		\exp
		\left\{
		\alpha_k\beta_k(x-x^\ast)
		\right\},
	\end{align*}
	where the second equality follows from
	\eqref{eq:mediator_contrast}. Hence, \linebreak
	$\operatorname{NIE}^{\log\lambda}_k(t;x,x^\ast)=\alpha_k\beta_k(x-x^\ast).$
	
	Equivalently, the natural indirect effect through~$M_k$ on the
	hazard-ratio scale is
	\[
	\operatorname{NIE}^{\mathrm{HR}}_k(t;x,x^\ast)=\exp\bigl\{\alpha_k\beta_k(x-x^\ast)\bigr\}.
	\]
	Because the proportional hazards model does not contain an
	exposure--mediator interaction, this quantity does not depend on time~$t$.
	For a one-unit exposure contrast, $x-x^\ast=1$, we obtain $\operatorname{NIE}^{\log\lambda}_k=\alpha_k\beta_k.$
	
	The same result holds on the log cumulative-hazard scale. Let $\Lambda_0(t)=\int_0^t\lambda_0(u)\,du.$ Under \eqref{eq:counterfactual_cox}, $\Lambda(t\mid x,\bm m,\bm Z)=\Lambda_0(t)\exp\{\gamma x+\bm\beta^\top\bm m+g(\bm Z)\}$.
	Therefore,
	\begin{align*}
		\operatorname{NIE}^{\log\Lambda}_k
		(t;x,x^\ast)
		&=
		\log
		\frac{
			\Lambda
			\left[
			t\mid x,\bm M^{(k)}(x,x^\ast),\bm Z
			\right]
		}{
			\Lambda
			\left[
			t\mid x,\bm M(x^\ast),\bm Z
			\right]
		}
		\nonumber\\
		&=
		\alpha_k\beta_k(x-x^\ast).
	\end{align*}
	Thus, $\alpha_k\beta_k$ has an exact causal interpretation as the
	path-specific indirect effect per unit change in $X$ on both the
	log-hazard and log cumulative-hazard scales, provided that the causal
	identification assumptions stated in Supplementary Section~S1.1 hold.
	
	We use deep neural networks (DNNs) as function approximators to capture the non-linear dependencies within the proposed causal mediation pathways. Specifically, the nuisance functions $g$ and $g_k$ are approximated by sparse feedforward networks with rectified linear unit (ReLU) activations \citep{glorot2011}, uniformly bounded weights and biases, and dropout regularization \citep{srivastava2014}. The formal definition of this function class, together with the network architecture conditions, is provided in Supplementary Section~S1.2.
	
	\subsection{Three-Step Estimation and Inference Procedure}
	Our objective is to identify components of the high-dimensional baseline damage spectrum that statistically mediate the association between sequence composition and storage failure. Because the number of candidate damage components can be large, direct simultaneous estimation and inference may be unstable and computationally demanding. We therefore develop a three-step procedure consisting of DNN-based mediator screening, SCAD-penalized estimation, and joint significance testing. This choice is motivated by
	the distinct roles of screening, sparse estimation, and inference in the proposed
	procedure. The first-stage screening is used only for dimension reduction and is designed to retain a moderately large set
	of candidate mediators. The retained mediators are subsequently fitted using SCAD penalization, which further removes irrelevant mediators. Finally, for each selected mediator, we test its significance.
	
	\paragraph{Step 1. DNN-Based Mediator Screening}
	\label{step1}
	For each candidate mediator \(M_k\), the screening step separately estimates the exposure-to-mediator coefficient \(\alpha_k\), following the DP2LM framework \citep{wang2024dp2lm}, and the mediator-to-outcome coefficient \(\beta_k\), using a deep partially linear Cox model \citep{zhong2022dplcox}. DNNs are used to flexibly approximate the nuisance functions associated with the covariates \(\mathbf Z\). Let \(\mathcal D_n\) denote the prespecified DNN function class. We estimate \(\alpha_k\) and \(g_k\) by
	\begin{equation*}
		(\hat{\alpha}_k,\hat{g}_k)=\arg\min_{\alpha_{k} \in \mathbb{R}, \, g_k \in \mathcal{D}_n} \,
		\frac{1}{n} \sum_{i=1}^n \{ M_{ik} - \alpha_{k} X_i - g_k(Z_i) \}^2.
	\end{equation*}
	We estimate the mediator-to-outcome association for each candidate mediator using a deep partially linear Cox model. Because this step is intended for dimension reduction, each mediator is considered separately rather than fitting the full high-dimensional mediator vector before screening. For the $k$th mediator, we specify
	\begin{equation*}
		\lambda_k(t\mid X_i,M_{ik},\mathbf Z_i)
		=
		\lambda_{0k}(t)
		\exp\left\{
		\gamma_k X_i+\beta_k M_{ik}+h_k(\mathbf Z_i)
		\right\},
	\end{equation*}
	where $h_k(\cdot)$ represents the potentially nonlinear effect of the covariates and is approximated by a DNN. The coefficient $\beta_k$ characterizes the conditional association between the $k$th damage component and the storage-failure hazard after adjustment for GC content and other sequence characteristics. For $k=1,\ldots,p$, we estimate $(\gamma_k,\beta_k,h_k)$ by maximizing the Cox partial log-likelihood:
	\begin{equation*}
		\begin{aligned}
			(\widehat\gamma_k,\widehat\beta_k,\widehat h_k)
			=
			\arg\max_{\substack{
					\gamma_k,\beta_k\in\mathbb R\\
					h_k\in\mathcal D
			}}
			\frac{1}{n}
			\sum_{i=1}^{n}
			\Delta_i
			\Bigg[
			&\gamma_kX_i+\beta_kM_{ik}+h_k(\mathbf Z_i)
			\\
			&\hspace{-9.8em}-
			\log\left\{
			\sum_{j:Y_j\ge Y_i}
			\exp\left(
			\gamma_kX_j+\beta_kM_{jk}+h_k(\mathbf Z_j)
			\right)
			\right\}
			\Bigg].
		\end{aligned}
	\end{equation*}
	Combining $\widehat\alpha_k$ from the mediator model and $\widehat\beta_k$ from the survival model, we define the product-of-coefficients screening score as $
	S_k
	=|
	\widehat\alpha_k\widehat\beta_k|,
	\qquad
	k=1,\ldots,p.$
	A large value of $S_k$ indicates that the $k$th damage component is associated with both the exposure and the storage-failure outcome, and hence is potentially important for the mediation pathway. We retain the top $d_n$ mediators based on their screening scores, where $
	\widehat{\mathcal B}
	=
	\{k: S_k \text{ belongs to the largest } d_n \text{ values}\},$ and $d_n = \min\left\{\left\lfloor \frac{n}{\log n} \right\rfloor, p\right\}.$ The threshold $d_n$ balances dimensionality reduction with the preservation of potentially important mediators.
	
	\paragraph{Step 2. SCAD-Penalized Estimation.}
	\label{step:scad}
	Let $\mathbf M_{i\widehat{\mathcal B}}$ denote the corresponding mediator
	subvector for the $i$th oligo. We jointly estimate the exposure effect,
	the mediator effects, and the nonlinear covariate component using the
	partially linear Cox model
	\begin{equation*}
		\lambda
		\left(
		t\mid X_i,\mathbf M_{i\widehat{\mathcal B}},\mathbf Z_i
		\right)
		=
		\lambda_0(t)
		\exp
		\left\{
		\gamma X_i
		+
		\boldsymbol{\beta}_{\widehat{\mathcal B}}^{\top}
		\mathbf M_{i\widehat{\mathcal B}}
		+
		g(\mathbf Z_i)
		\right\}.
	\end{equation*}
	The corresponding empirical log partial likelihood is
	\begin{equation*}
		\begin{aligned}
			\ell_n
			\left(
			\gamma,
			\boldsymbol{\beta}_{\widehat{\mathcal B}},
			g
			\right)
			=
			\frac{1}{n}
			\sum_{i=1}^{n}
			\Delta_i
			\Bigg[
			&\gamma X_i
			+
			\boldsymbol{\beta}_{\widehat{\mathcal B}}^{\top}
			\mathbf M_{i\widehat{\mathcal B}}
			+
			g(\mathbf Z_i)
			\\
			&\hspace{-9.5em}-
			\log
			\left\{
			\sum_{j:Y_j\geq Y_i}
			\exp
			\left(
			\gamma X_j
			+
			\boldsymbol{\beta}_{\widehat{\mathcal B}}^{\top}
			\mathbf M_{j\widehat{\mathcal B}}
			+
			g(\mathbf Z_j)
			\right)
			\right\}
			\Bigg].
		\end{aligned}
	\end{equation*}
	
	For $u>0$, the derivative of the SCAD penalty is defined as
	\begin{equation*}
		p_{\lambda}'(u)
		=
		\lambda I(u\leq\lambda)
		+
		\frac{(a\lambda-u)_{+}}{a-1}
		I(u>\lambda),
	\end{equation*}
	where $\lambda>0$ controls the overall level of penalization,
	$a>2$ is the shape parameter, conventionally set to $a=3.7$ \citep{fan2001},
	$I(\cdot)$ denotes the indicator function, and
	$(x)_{+}=\max(x,0)$.
	
	We define the penalized objective function \citep{sun2024dplc} as:
	\begin{equation*}
		Q_n
		\left(
		\gamma,
		\boldsymbol{\beta}_{\widehat{\mathcal B}},
		g
		\right)
		=
		-
		\ell_n
		\left(
		\gamma,
		\boldsymbol{\beta}_{\widehat{\mathcal B}},
		g
		\right)
		+
		\sum_{k\in\widehat{\mathcal B}}
		p_{\lambda}
		\left(
		|\beta_k|
		\right).
	\end{equation*}
	The SCAD-penalized estimator is given by
	\begin{equation*}
		\left(
		\widehat{\gamma},
		\widehat{\boldsymbol{\beta}}_{\widehat{\mathcal B}},
		\widehat g
		\right)
		=
		\arg\min_{\substack{
				\gamma\in\mathbb R,\,
				\boldsymbol{\beta}_{\widehat{\mathcal B}}
				\in\mathbb R^{|\widehat{\mathcal B}|}\\
				g\in\mathcal D
		}}
		Q_n
		\left(
		\gamma,
		\boldsymbol{\beta}_{\widehat{\mathcal B}},
		g
		\right).
	\end{equation*}
	The set of mediators selected by the SCAD procedure is $
	\widehat{\mathcal A}
	=
	\left\{
	k\in\widehat{\mathcal B}:
	\widehat{\beta}_k\neq 0
	\right\}.$
	
	\paragraph{Step 3. Joint Significance Testing.} For each
	$k\in\widehat{\mathcal A}$, we assess whether the $k$th damage
	component exhibits a statistically significant mediation signal along
	the pathway $
	X\longrightarrow M_k\longrightarrow T.$ The corresponding composite null hypothesis is $H_{0k}:\alpha_k\beta_k=0,$ which can be written as the union of two component null hypotheses, $H_{0k}^{(\alpha)}:\alpha_k=0
	\qquad\text{or}\qquad
	H_{0k}^{(\beta)}:\beta_k=0.$ Accordingly, evidence for mediation through $M_k$ requires evidence
	against both component null hypotheses.
	
	For the exposure-to-mediator pathway, we construct the Wald statistic $Z_k^{(\alpha)}
	=
	\frac{\widehat{\alpha}_k}
	{\widehat{\operatorname{se}}(\widehat{\alpha}_k)},$ where $\widehat{\alpha}_k$ is the estimated exposure-to-mediator coefficient
	from Step~\ref{step1}, and $\widehat{\operatorname{se}}(\widehat{\alpha}_k)$ is the
	standard error from ordinary least squares regression after DNN-based
	confounding adjustment \citep{wang2024dp2lm}. The corresponding two-sided $p$-value is defined as
	\begin{equation*}
		P_k^{(\alpha)}
		=
		2\left[
		1-\Phi\left(
		\left|Z_k^{(\alpha)}\right|
		\right)
		\right]
		=
		2\left[
		1-\Phi\left(
		\frac{|\widehat{\alpha}_k|}
		{\widehat{\operatorname{se}}(\widehat{\alpha}_k)}
		\right)
		\right].
	\end{equation*}
	
	We define $Z_k^{(\beta)}
	=
	\frac{\widetilde{\beta}_k}
	{\widehat{\operatorname{se}}(\widetilde{\beta}_k)},$ where $\widetilde{\beta}_k$ is the unpenalized refit estimate obtained at the end of Step~\ref{step:scad}, and $\widehat{\operatorname{se}}(\widetilde{\beta}_k)$ is obtained from the semiparametric efficient information matrix for the partially linear Cox model, estimated by projecting $(X,\mathbf{M}_{\widehat{\mathcal{A}}})$ onto $(Y,\mathbf{Z})$ conditional on observed events ($\Delta_i=1$) using a deep network \citep{zhong2022dplcox}.
	Its two-sided $p$-value is
	\begin{equation*}
		P_k^{(\beta)}
		=
		2\left[
		1-\Phi\left(
		\left|Z_k^{(\beta)}\right|
		\right)
		\right]
		=
		2\left[
		1-\Phi\left(
		\frac{|\widetilde{\beta}_k|}
		{\widehat{\operatorname{se}}(\widetilde{\beta}_k)}
		\right)
		\right].
	\end{equation*}
	Here, $\Phi(\cdot)$ denotes the cumulative distribution function of the
	standard normal distribution.
	
	Because the composite null hypothesis $H_{0k}$ is rejected only when
	both component null hypotheses are rejected, the joint-significance
	$p$-value for mediator $M_k$ is defined by the intersection--union
	principle as $P_k^{\mathrm{JS}}
	=
	\max
	\left\{
	P_k^{(\alpha)},
	P_k^{(\beta)}
	\right\}.$ Thus, a small value of $P_k^{\mathrm{JS}}$ requires simultaneous
	evidence that the exposure is associated with the $k$th mediator and
	that the mediator is associated with the storage-failure hazard after
	adjustment for the exposure and other covariates.
	
	To account for multiplicity among the mediators retained for inference,
	the joint-significance $p$-values are adjusted according to $P_{k,\mathrm{adj}}^{\mathrm{JS}}
	=
	\min
	\left\{
	|\widehat{\mathcal A}|\,
	P_k^{\mathrm{JS}},
	1
	\right\},
	\qquad
	k\in\widehat{\mathcal A}.$ At a prespecified significance level $\delta$, the $k$th mediator is
	declared statistically significant if
	$P_{k,\mathrm{adj}}^{\mathrm{JS}}<\delta.$ In the numerical studies and the DNA storage application, we take $\delta=0.05$.
	
	\section{Simulation Studies}
	\label{sec:simulations}
	In this section, we conduct simulation studies to evaluate the finite-sample performance of the proposed method under both general and DNA-storage-inspired settings. To mirror the DNA storage application, we let $X$ represent a standardized sequence-composition feature, such as centered and scaled GC content. The covariate vector $\mathbf{Z}$ represents additional sequence characteristics, $\mathbf{M}$ represents a high-dimensional baseline damage spectrum, and the scalar event time $T$ represents time to storage-quality failure. Note that the raw GC proportion is naturally bounded between 0 and 1; here $X$ is interpreted as its standardized counterpart rather than the raw GC proportion. The simulations are designed to examine whether the proposed procedure can identify sparse damage-mediated pathways under nonlinear covariate effects. We further consider different covariate distributions to assess robustness to departures from specific distributional assumptions. For \(i = 1, 2, \ldots, n\), the exposure \(X_i\) and the covariate vector \(\mathbf{Z}_i = (Z_{i1}, Z_{i2}, Z_{i3}, Z_{i4}, Z_{i5}, Z_{i6})^{\top}\) are generated from one of five distributional settings:
	\[
	\begin{aligned}
		\text{Normal:}\quad
		&X_i\sim N(0,1),
		&&Z_{ij}\sim N(0,1);\\
		\text{Laplace:}\quad
		&X_i\sim\operatorname{Laplace}(0,0.5),
		&&Z_{ij}\sim\operatorname{Laplace}(0,0.5);\\
		\text{Uniform:}\quad
		&X_i\sim U(-1,1),
		&&Z_{ij}\sim U(-1,1);\\
		\text{Normal--Uniform:}\quad
		&X_i\sim N(0,1),
		&&Z_{ij}\sim U(-1,1);\\
		\text{Uniform--Laplace:}\quad
		&X_i\sim U(-1,1),
		&&Z_{ij}\sim\operatorname{Laplace}(0,0.5).
	\end{aligned}
	\]
	Let $\gamma=0.5$; $\beta_1 = 0.6$, $\beta_2 = -0.5$, $\beta_3 = 0.4$, $\beta_4 = -0.55$, $\beta_5 = 0.45$, $\beta_6 = 0.7$, $\beta_7 = -0.45$, $\beta_8 = 0.5$, $\beta_9 = 0.75$, $\beta_{10} = 0.65$, and $\beta_k = 0$ otherwise; $\alpha_1 = 0.6$, $\alpha_2 = -0.5$, $\alpha_3 = 0.4$, $\alpha_4 = -0.55$, $\alpha_5 = 0.45$, $\alpha_6 = 0.7$, $\alpha_7 = -0.45$, $\alpha_8 = 0.5$, $\alpha_9 = 0.75$, $\alpha_{10} = 0.6$, and $\alpha_k = 0$ otherwise, i.e.\ $\{M_k\}_{k=1}^{10}$ are the active mediators generated from the mediator model. The error term $\varepsilon_i$ follows a multivariate Gaussian distribution, $\mathcal{N}_p(\mathbf{0}, \boldsymbol{\Sigma})$, where the covariance matrix $\boldsymbol{\Sigma}$ exhibits a first-order autoregressive structure. Specifically, the covariance between the $j$-th and $k$-th elements is defined as $\mathrm{Cov}(\varepsilon_{i,j}, \varepsilon_{i,k}) = \rho^{|j-k|}$ with $\rho = 0.25$. Here, $g_{k}(\mathbf{z})=0.1z_{1}z_{2}+0.2\cos(\pi z_{3}z_{4})z_{5}+(z_{6}+5)^{-1}$ and $g(\mathbf{z})=\frac{1}{8}\left[z_{1}^{2}z_{2}^{3}+\log(z_{3}^2)+\sqrt[3]{z_{4}z_{5}}+\exp\left(\frac{z_{5}}{2}\right)\right]-3$. The survival time $T_i$ is generated from a Cox proportional hazards model with an exponential baseline hazard $\lambda_0(t)=1$. Under this baseline, the cumulative hazard is $\Lambda_0(t)=t$, and the survival time can be generated via inverse-transform sampling. Specifically, given the linear predictor $\eta_i = \gamma X_i + \sum_{k=1}^{p} \beta_k M_{ik} + g(\mathbf{Z}_i),$
	we generate $U_i \sim \text{Uniform}(0,1)$ independently and set $T_i = -\log(1-U_i) \exp(-\eta_i).$
	This construction ensures that $T_i$ follows the specified Cox model with hazard function $\lambda(t \mid X_i, \mathbf{M}_i, \mathbf{Z}_i) = \lambda_0(t) \exp(\eta_i)$. The censoring time $C_i$ was simulated from a uniform distribution on $[0,c]$, where $c$ was chosen so that the censoring rate was approximately 20\% and 40\%.	
	
	We conducted simulations with varying sample sizes. Specifically, the
	total number of mediators was fixed at $p=300$, among which 10 were
	truly active, while the sample size was varied over
	$n\in\{500,1000\}$. The performance of each method was evaluated over
	200 replications. For comparison, we included
	\textit{survHIMA}~\citep{zhang2021}, as a benchmark. To ensure a fair comparison, the same simulated datasets
	and data-generating mechanisms were used for both the proposed method
	and \textit{survHIMA} in each replication.
	
	We tuned the hyperparameters for each estimation step using the simulated datasets with 80\%-20\% training-validation splits. For the neural network architecture in the mediator screening stage (Step~\ref{step1}), we used a two-hidden-layer configuration with 16 and 8 neurons, respectively, for the exposure-to-mediator pathway, and a two-hidden-layer configuration with 100 neurons per layer for the mediator-to-outcome pathway. For the SCAD estimation stage (Step~\ref{step:scad}), we used an 8-8 architecture for the nonlinear confounding adjustment. The dropout rate was set to 0.1 for Step 1 and 0.5 for Step~\ref{step:scad}. All networks were trained using the Adam optimizer with full-batch gradient descent. Early stopping with a patience of 20 epochs was applied for the mediator pathway networks based on validation loss. For the SCAD penalty in Step~\ref{step:scad}, we selected the optimal tuning parameter $\lambda$ from a candidate sequence in $[0.01, 0.5]$ based on the extended Bayesian information criterion (EBIC) \citep{chen2008}.

	\begin{figure*}[t]
		\centering
		\includegraphics[width=\linewidth]{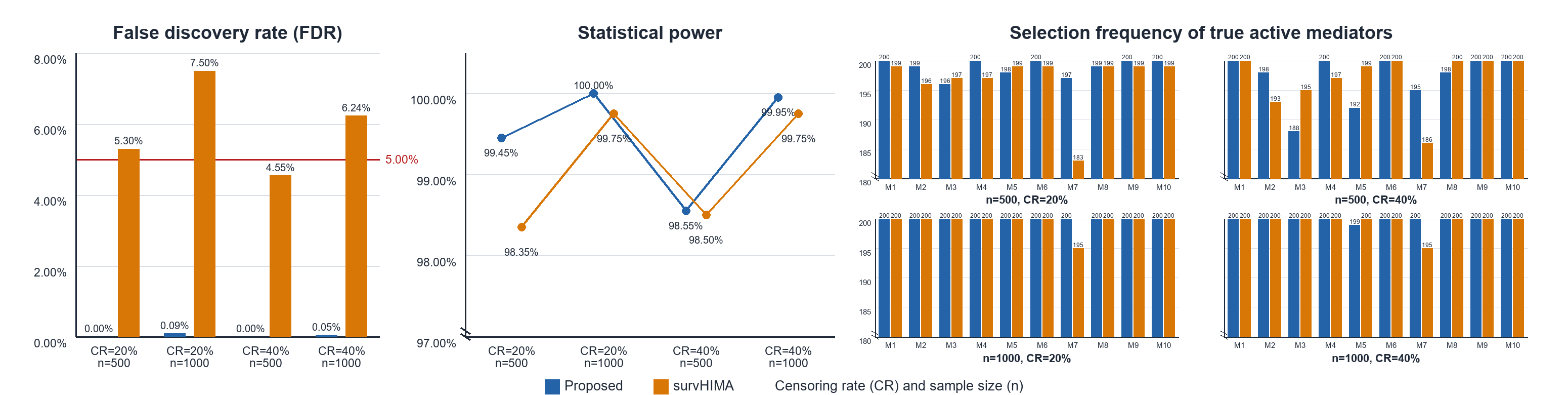}
		\caption{False discovery rate and statistical power under the normal-distribution setting.}
		\label{fig:simulation-fdr-power}
		\figalttext[FDR, power plots and the frequency of selection under normal distribution]{Four-panel plot showing false discovery rate and statistical power for proposed method versus survHIMA at two censoring rates (20\% and 40\%) and two sample sizes (n=500 and n=1000) under normal distribution setting.}
	\end{figure*}
	
	\begin{table}[H]
		\centering
		\footnotesize
		\setlength{\tabcolsep}{3pt}
		\caption{Bias and MSE (in parentheses) of the estimated mediation effects under the normal-distribution setting}
		\label{tab:normal-bias-mse}
		\begin{tabular}{ccccccccc}
			\toprule
			& \multicolumn{4}{c}{CR = 20\%} & \multicolumn{4}{c}{CR = 40\%} \\
			\cmidrule(lr){2-5} \cmidrule(lr){6-9}
			& \multicolumn{2}{c}{n = 500} & \multicolumn{2}{c}{n = 1000} 
			& \multicolumn{2}{c}{n = 500} & \multicolumn{2}{c}{n = 1000} \\
			\cmidrule(lr){2-3} \cmidrule(lr){4-5} 
			\cmidrule(lr){6-7} \cmidrule(lr){8-9}
			& Proposed & survHIMA & Proposed & survHIMA & Proposed & survHIMA & Proposed & survHIMA \\
			\midrule
			\multirow{2}{*}{$\alpha_1 \beta_1$} 
			& -0.0222 & -0.0713 & -0.0438 & -0.0749 & -0.0238 & -0.0685 & -0.0424 & -0.0669 \\
			& (0.0040) & (0.0073) & (0.0035) & (0.0071) & (0.0046) & (0.0073) & (0.0035) & (0.0058) \\
			\cmidrule(lr){1-9}
			\multirow{2}{*}{$\alpha_2 \beta_2$} 
			& -0.0164 & -0.0503 & -0.0345 & -0.0552 & -0.0203 & -0.0512 & -0.0335 & -0.0496 \\
			& (0.0030) & (0.0043) & (0.0021) & (0.0040) & (0.0037) & (0.0044) & (0.0022) & (0.0034) \\
			\cmidrule(lr){1-9}
			\multirow{2}{*}{$\alpha_3 \beta_3$} 
			& -0.0157 & -0.0386 & -0.0195 & -0.0335 & -0.0177 & -0.0390 & -0.0186 & -0.0299 \\
			& (0.0017) & (0.0024) & (0.0010) & (0.0017) & (0.0024) & (0.0025) & (0.0011) & (0.0015) \\
			\cmidrule(lr){1-9}
			\multirow{2}{*}{$\alpha_4 \beta_4$} 
			& -0.0223 & -0.0660 & -0.0322 & -0.0587 & -0.0223 & -0.0609 & -0.0308 & -0.0524 \\
			& (0.0034) & (0.0064) & (0.0022) & (0.0047) & (0.0037) & (0.0058) & (0.0022) & (0.0039) \\
			\cmidrule(lr){1-9}
			\multirow{2}{*}{$\alpha_5 \beta_5$} 
			& -0.0146 & -0.0440 & -0.0248 & -0.0422 & -0.0162 & -0.0398 & -0.0218 & -0.0370 \\
			& (0.0024) & (0.0035) & (0.0015) & (0.0026) & (0.0034) & (0.0033) & (0.0015) & (0.0023) \\
			\cmidrule(lr){1-9}
			\multirow{2}{*}{$\alpha_6 \beta_6$} 
			& -0.0228 & -0.0932 & -0.0523 & -0.0936 & -0.0206 & -0.0826 & -0.0498 & -0.0817 \\
			& (0.0057) & (0.0126) & (0.0049) & (0.0112) & (0.0064) & (0.0110) & (0.0053) & (0.0089) \\
			\cmidrule(lr){1-9}
			\multirow{2}{*}{$\alpha_7 \beta_7$} 
			& -0.0152 & -0.0390 & -0.0229 & -0.0401 & -0.0148 & -0.0357 & -0.0229 & -0.0354 \\
			& (0.0021) & (0.0027) & (0.0014) & (0.0024) & (0.0024) & (0.0026) & (0.0016) & (0.0020) \\
			\cmidrule(lr){1-9}
			\multirow{2}{*}{$\alpha_8 \beta_8$} 
			& -0.0182 & -0.0559 & -0.0261 & -0.0502 & -0.0139 & -0.0486 & -0.0259 & -0.0447 \\
			& (0.0025) & (0.0048) & (0.0019) & (0.0037) & (0.0027) & (0.0041) & (0.0021) & (0.0031) \\
			\cmidrule(lr){1-9}
			\multirow{2}{*}{$\alpha_9 \beta_9$} 
			& -0.0281 & -0.0943 & -0.0633 & -0.1049 & -0.0270 & -0.0826 & -0.0565 & -0.0873 \\
			& (0.0075) & (0.0138) & (0.0071) & (0.0136) & (0.0088) & (0.0117) & (0.0067) & (0.0103) \\
			\cmidrule(lr){1-9}
			\multirow{2}{*}{$\alpha_{10} \beta_{10}$} 
			& -0.0152 & -0.0711 & -0.0461 & -0.0793 & -0.0102 & -0.0633 & -0.0415 & -0.0694 \\
			& (0.0037) & (0.0075) & (0.0038) & (0.0079) & (0.0039) & (0.0063) & (0.0034) & (0.0063) \\
			\bottomrule
		\end{tabular}
	\end{table}
	
	We compare the proposed method with \textit{survHIMA}~\citep{zhang2021}.
	For conciseness, the main text reports the normal-distribution results in Figure~\ref{fig:simulation-fdr-power} and Table~\ref{tab:normal-bias-mse}. Results for the remaining distributional settings are provided in Supplementary Tables S1--S12 to further assess the robustness of the proposed method.
	
	\paragraph{Statistical power.}
	As shown in Figure~\ref{fig:simulation-fdr-power}, the proposed method demonstrated consistently high screening performance for all ten active mediators, reaching near-perfect screening at $n=1000$. Increasing the censoring rate from 20\% to 40\% generally reduced power for both approaches, as fewer observed failures provide less information for estimating the survival coefficients $\beta$. For the proposed method, the overall power (averaged across all ten mediators) decreased from 99.45\% to 98.55\% at $n=500$ (Normal distribution) and from 100\% to 99.95\% at $n=1000$, remaining above 89.5\% in every distributional setting even at the higher censoring rate. \textit{survHIMA} showed comparable power across censoring rates, at 98.35\% versus 98.50\% for $n=500$ and 99.75\% versus 99.75\% for $n=1000$. The largest reductions occurred for mediators \(M_3\) and \(M_7\), which have weaker effects. Notably, \textit{survHIMA} exhibited more pronounced degradation for these weaker mediators, particularly \(M_7\). 
	
	\paragraph{False discovery rate control.}
	Figure~\ref{fig:simulation-fdr-power} and Supplementary Tables S2, S5, S8, and S11 show that the proposed method provided strict control of false discoveries. The observed FDR ranged from 0\% to 0.14\% across all combinations of distribution, sample size, and censoring rate, remaining well below the nominal 5\% level. In contrast, \textit{survHIMA} exhibited higher false discovery rates, ranging from 3.48\% to 7.50\%, occasionally exceeding the 5\% threshold. This difference is particularly pronounced at $n=1000$ under the Normal distribution, where \textit{survHIMA}'s FDR reached 7.5\% (20\% censoring) and 6.24\% (40\% censoring), while the proposed method maintained FDR below 0.1\%.
	
	\paragraph{Bias and mean squared error (MSE) reduction.}
	Table~\ref{tab:normal-bias-mse} and Supplementary Tables S3, S6, S9, and S12 show that the proposed method substantially reduced both bias and MSE relative to \textit{survHIMA}. The average absolute bias dropped by approximately 72.2\% across all settings. For example, under the Normal distribution with \(n=500\) and 20\% censoring, the bias for \(\alpha_1\beta_1\) was \(-0.0222\) (proposed) versus \(-0.0713\) (\textit{survHIMA}), a 68.9\% reduction; for \(\alpha_9\beta_9\), it was \(-0.0281\) versus \(-0.0943\) (70.2\% reduction). MSE decreased by 9.7\% on average, with larger gains under 20\% and 40\% censoring. These results suggest that the DNN-based nonlinear method captures covariate effects more accurately than the linear specification, particularly for mediators with larger effect sizes (\(M_6\), \(M_9\), \(M_{10}\)).
	
	\paragraph{Implications for DNA storage reliability.}
	From the perspective of DNA storage reliability, these simulation results indicate that the proposed procedure can distinguish potentially important sequence-dependent damage mechanisms from a high-dimensional baseline damage spectrum, identifying 8--10 out of 10 true mediators with near-zero false positives.
	
	\section{DNA Data Storage Application}
	\label{sec:application}
	\subsection{Data Extraction}
	\label{subsec:4.1}
\begin{figure}[H]
		\centering
		\includegraphics[width=\linewidth]{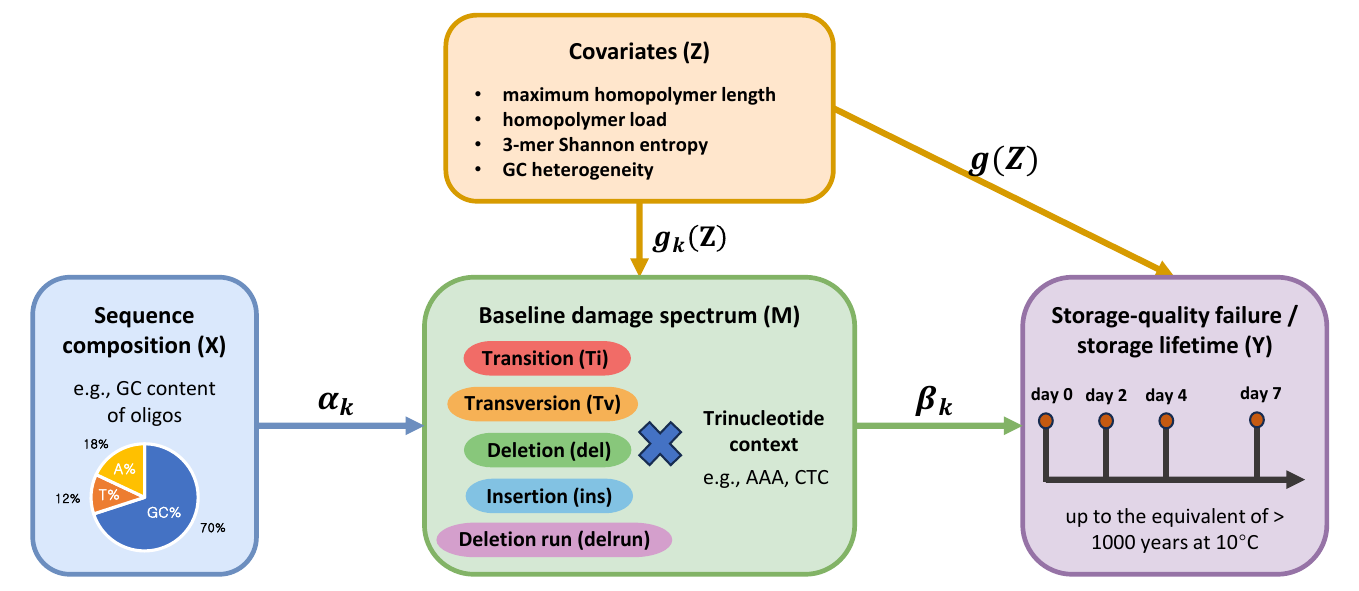}
		\caption{Experimental design and data structure for the
			k-th Mediator. The aging experiment employed accelerated aging at 70\,$^{\circ}$C, with sequencing performed at multiple time points (0, 2, 4, and 7 days). Based on \citet{gimpel2023}, this protocol is equivalent to more than 1000 years of storage at 10\,$^{\circ}$C. }
		\label{fig:workflow}
	\end{figure}

	We apply the proposed framework to an aging experiment on synthetic DNA to investigate whether sequence composition affects archival reliability through specific components of the baseline damage spectrum. In this analysis, GC content serves as the exposure, context-specific sequencing errors observed at baseline form the high-dimensional mediator, and subsequent deterioration in read quality defines the storage outcome. 	Figure~\ref{fig:workflow} summarizes the experimental design and the construction of the exposure, mediator, and outcome data. The data come from an aging experiment on the Genscript pool in the digital twin study \citep{gimpel2023} and are publicly available at \url{https://www.ebi.ac.uk/ena/browser/view/PRJEB65931}.
	
	The data extraction proceeded in three stages, which are summarized schematically in Supplementary Figure S1. Starting from the design oligonucleotide pool ($N = 12{,}472$ sequences, 102 nt), we extract GC content as the exposure variable $X$, calculated as the proportion of G and C bases in each oligo, and covariates $Z$, including the maximum homopolymer length, homopolymer load, 3-mer Shannon entropy, and GC heterogeneity. Detailed definitions are provided in Supplementary Section~S3.
	
	We first applied coverage-based quality control to ensure adequate statistical power and measurement reliability. This procedure retained sequences with \(\geq 20\) reads at day 0. During follow-up (days 2, 4, 7), sequences were retained only if they had \(\geq 30\) reads or completely dropped out (coverage \(= 0\), recorded as failure events), yielding \(N = 6{,}803\) sequences (\(54.5\%\) of the original pool). To reduce dependence arising from shared
	reads and to limit shared measurement error between the mediator
	and outcome measurements, for each retained oligo, the day-0 reads were then randomly split into two non-overlapping halves of approximately equal size: the A-half was used to compute the 320-dimensional mediator $M$ (damage spectrum), and the B-half was reserved for defining baseline quality failure and population allocation.
	
	\paragraph{The definition of mediators} The 320-dimensional mediator $\mathbf{M}$ represents the baseline damage spectrum (day 0), comprising 64 trinucleotide contexts and 5 error types (Ti, Tv, del, ins, and delrun). Let $\mathcal{R}_{i,0}^{A}$ denote the set of usable A-half reads assigned to oligo $i$ at day 0. For each eligible oligo $i$, the mediator is defined as:
	\[
	M_{i,(c,k)} =
	\frac{
		\#\{\text{errors of type }k\text{ in context }c\text{ on oligo }i\}
	}{
		\#\{\mathcal{R}_{i,0}^{A}\}
	}.
	\]
	Here, $c$ denotes one of the 64 trinucleotide contexts, $k$ denotes one
	of the 5 error types, and $\#\{\cdot\}$ denotes the cardinality of a set.
	
	\paragraph{The definition of storage-quality failure} We construct an operational storage-quality failure outcome from
	sequencing read-quality data. Rather than attempting to measure the
	physical lifetime of individual DNA molecules, the proposed outcome
	characterizes deterioration in oligo-level retrievability based on the
	fraction of low-quality sequencing reads. All quantities used to
	construct the outcome are calculated from the B-half reads.
	
	Let
	$\mathcal{R}_{i,t}^{B}$ denote the set of usable B-half reads assigned to
	oligo $i$ at assessment time $t$. Let $e_i(r)$ be the Levenshtein edit distance \citep{Levenshtein1965} between read $r$ and
	the reference sequence of oligo $i$, let $L_i$ be the reference length,
	and let $\tau$ be the prespecified read-level quality threshold. Define the low-quality read fraction by
	\[
	f_i^{B}(t)
	=
	\frac{\#\left\{
		r\in\mathcal{R}_{i,t}^{B}: e_i(r)/L_i>\tau
		\right\}}{\#\{\mathcal{R}_{i,t}^{B}\}}.
	\]
	\paragraph{Part 1: Baseline quality failure.}
	We first identify oligos that already exhibit poor read quality at
	baseline. Define $B_i
	=
	I\left\{f_i^{B}(0)>\phi\right\},$ where $\phi$ is the oligo-level failure threshold. An oligo with
	$B_i=1$ is classified as a baseline quality failure, whereas an oligo
	with $B_i=0$ passes baseline quality control. Baseline quality failure
	is evaluated for the entire oligo pool. Only oligos satisfying
	$B_i=0$ enter the subsequent storage-lifetime analysis.
	
	\paragraph{Part 2: Threshold-defined storage-quality lifetime.}
	Among oligos that pass baseline quality control ($B_i=0$),
	storage-quality lifetime is operationally defined by the first
	post-baseline assessment at which the proportion of low-quality reads
	exceeds $\phi$. Let $T_i^{*}
	=
	\min
	\left\{
	t\in\{2,4,7\}:
	f_i^{B}(t)>\phi
	\right\},
	\qquad
	\min\varnothing=\infty.$ Because follow-up ends at day 7, the observed survival outcome is $
	Y_i=\min(T_i^{*},7),
	\qquad
	\Delta_i=I(T_i^{*}\le 7).$ Thus, $\Delta_i=1$ indicates that storage-quality failure is observed
	at one of the follow-up assessments, whereas $\Delta_i=0$ indicates
	administrative censoring at day 7. The resulting $Y_i$ should be
	interpreted as an operational, threshold-defined storage-quality
	lifetime rather than the physical degradation time of an individual
	DNA molecule.
	
	Prior to quality control, reads from each sequencing library were subsampled to 1{,}000{,}000 reads to standardize sequencing depth across time points. This yielded a median per-oligo coverage of 112 reads, which is sufficient for reliable error-rate estimation and damage spectrum construction.
	
	To operationalize the two-part storage-quality outcome defined above, we calibrated the read-level threshold $\tau$ and oligo-level threshold $\phi$ using a grid search. Specifically, the candidate values were
	\[
	\begin{aligned}
		\tau &\in \{0.01, 0.03, 0.05, 0.07, 0.10\},\\
		\phi &\in \{0.02, 0.025, 0.03, 0.05, 0.10, 0.15, 0.20, 0.25, 0.30, 0.40, 0.50\}.
	\end{aligned}
	\]
	The choice of $\tau$ is motivated by error levels that can be tolerated by representative DNA storage coding schemes. For reference, HEDGES achieves error-free recovery at error rates of approximately 1\%, 3\%, and 7--10\% for code rates of 0.6, 0.5, and 0.25, respectively \citep{press2020}. We use $\tau=0.03$ and $\phi=0.3$ as the analysis thresholds. Sequences with $f_i^{B}(0) > \phi$ ($n = 613$, 4.9\%) exhibited baseline quality failure. The remaining $n = 6{,}190$ sequences (49.6\% of the original pool) constitute the analysis sample. The overall censoring rate was 77.2\%. Alternative values are evaluated in sensitivity analyses, with results provided in Supplementary Section~S4.
	
	\subsection{Identification of Damage-Mediated Pathways}
	We applied the proposed partially linear mediation model to investigate the
	pathways linking GC content to storage-quality failure through the
	high-dimensional baseline damage spectrum, while adjusting for the
	sequence-derived covariates $\mathbf{Z}$. The SCAD-penalized procedure
	selected 22 mediators, of which 14 passed the joint
	significance test at the 0.05 level. The estimated coefficient of GC content was $\widehat{\gamma}=0.253.$
	
	For each selected mediator $M_k$, the estimated mediated pathway effect was
	quantified by $\widehat{\alpha}_k\hat{\beta}_k$, consistent with the
	mediation-effect definition introduced in Section~\ref{subsec:causal_interpretation}. The sum of the
	estimated pathway effects over the 22 SCAD-selected mediators was $
	\sum_{k\in\widehat{\mathcal A}}
	\widehat{\alpha}_k\hat{\beta}_k
	=0.166.$
	
\begin{figure}[H]
		\centering
		\includegraphics[width=\linewidth]{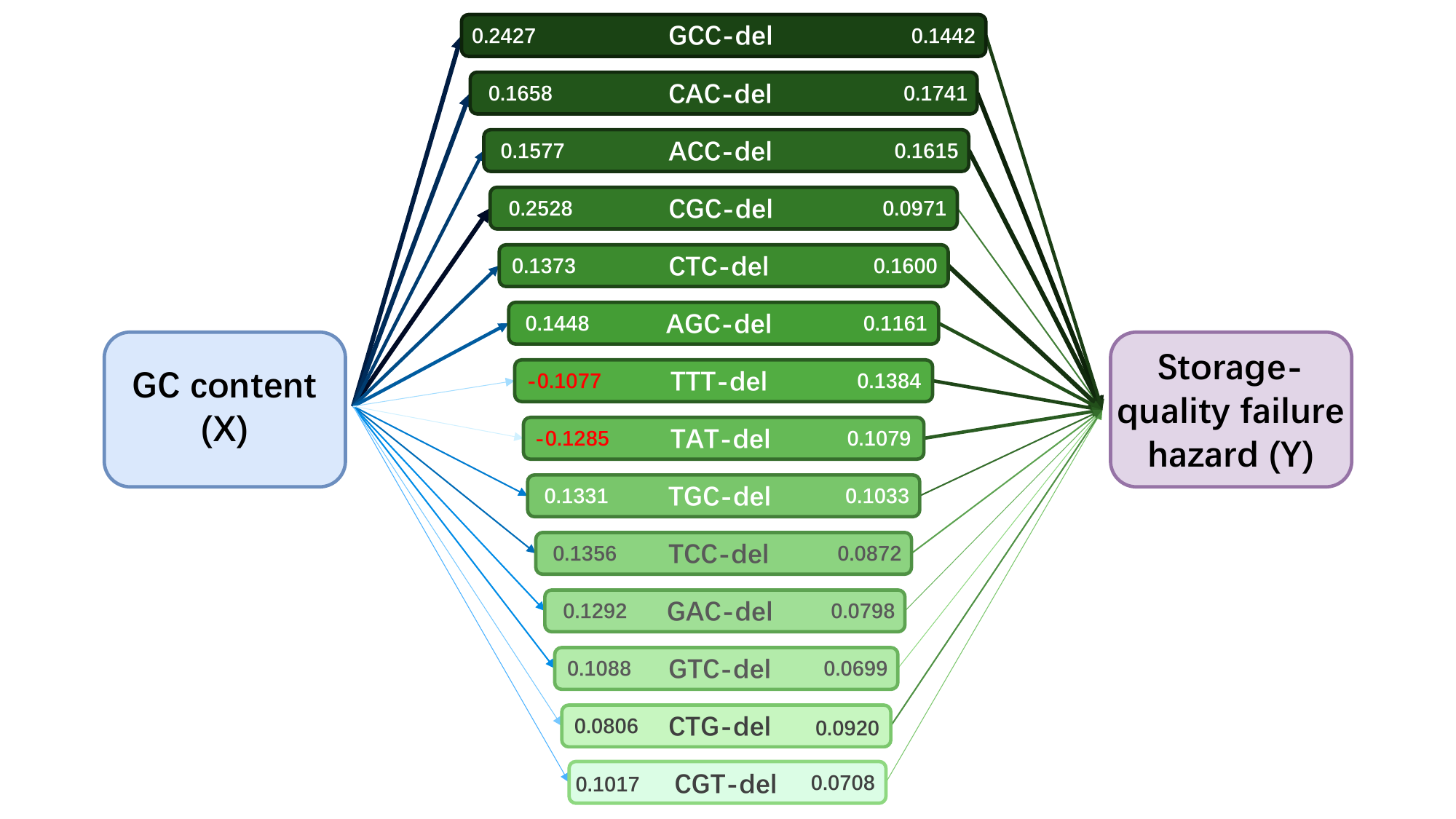}
		\caption{Significant mediation pathways from GC content to storage-quality failure. Arrow width and color intensity are proportional to the mediated effect $|\hat{\alpha}_k\hat{\beta}_k|$; numbers adjacent to arrows denote the exposure--mediator effect $\hat{\alpha}_k$ and the mediator--outcome effect $\hat{\beta}_k$. Only pathways with $p_{\text{joint}} < 0.05$ are shown.}
		\label{fig:app_pathway}
	\end{figure}

	For the Genscript pool at $\tau=0.03$, $\phi=0.30$, the test-set C-index was $0.857$~\citep{Harrell1996}. Figure~\ref{fig:app_pathway} displays the 14 significant mediators and their estimated pathway coefficients. Remarkably, all 14 significant mediators correspond to context-specific single-base deletions. Among the 14 pathways, 12 exhibit positive mediated effects $\widehat{\alpha}_k \hat{\beta}_k > 0$, indicating that higher GC content is associated with increased baseline damage burden in these contexts, which in turn is associated with elevated failure hazard. The remaining two pathways (TTT-del and TAT-del) show negative mediated effects. For these mediators, the product $\widehat{\alpha}_k \hat{\beta}_k < 0$ arises because the exposure-to-mediator coefficient is negative whereas the mediator-to-outcome coefficient is positive, resulting in a suppressing or inconsistent mediation pattern. These negative pathways should not be interpreted as protective; rather, they represent conditional associations after adjustment for GC content, other sequence characteristics, and the remaining selected damage components.
	
	Among the positive pathways, 10 out of 12 occur in trinucleotide contexts ending in cytosine (C). This enrichment suggests that the deletion-mediated pathway linking GC content to storage failure is concentrated in specific sequence contexts.
	Specifically, \(\alpha_k>0\) indicates that higher GC content is associated with a greater burden of the \(k\)th damage component, whereas \(\beta_k>0\) indicates that a greater burden of that component is associated with a higher failure hazard. When both coefficients are positive, a positive product \(\alpha_k\beta_k\) therefore represents a pathway through which higher GC content is associated with increased failure risk via increased baseline damage.
	
	The predominance of deletion-mediated pathways is consistent with known properties of DNA synthesis and error correction. Insertions and deletions are particularly challenging for many coding schemes because they disrupt sequence synchronization \citep{press2020}. In electrochemical DNA synthesis, deletions can arise from mass transfer limitations as oligonucleotides grow longer: the distance to the acid-generating electrode increases, and steric hindrance impedes acid-induced deprotection, preventing addition of the next nucleotide \citep{gimpel2023}. Deletions also tend to cluster within reads rather than being distributed independently \citep{gimpel2023}. 
	
\begin{figure}[H]
		\centering
		\includegraphics[width=\linewidth]{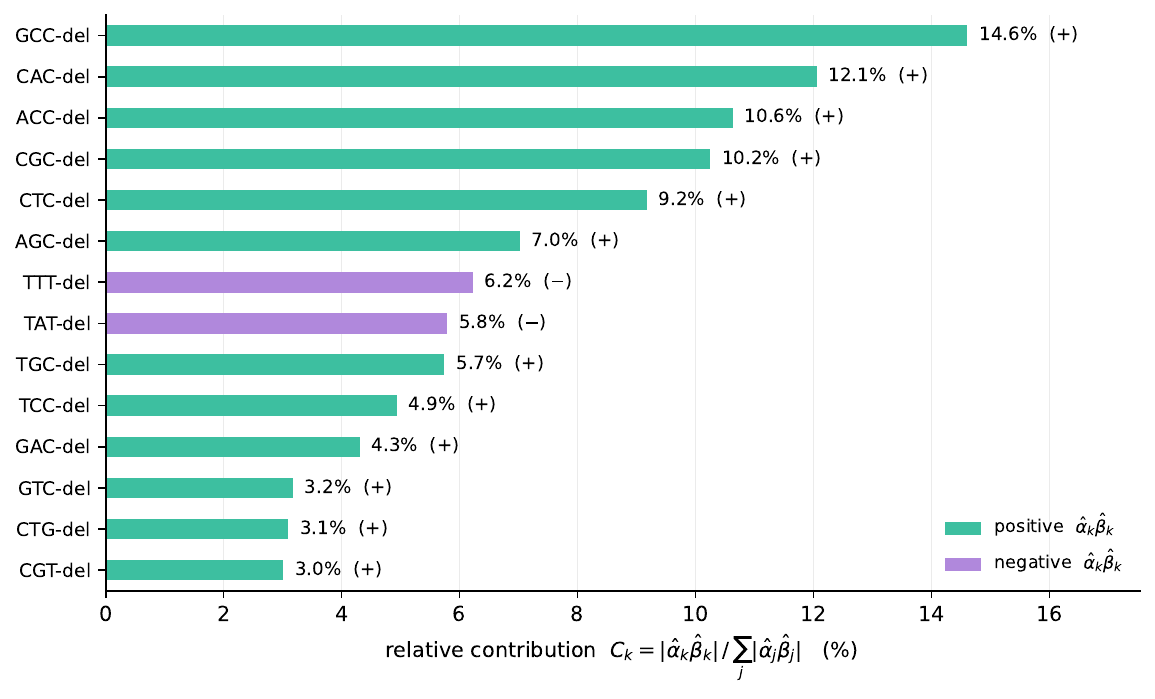}
		\caption{Relative magnitudes and directions of the significant damage-mediated pathways among all candidate mediators.}
		\label{fig:contrib}
	\end{figure}

	The enrichment of C-ending trinucleotide contexts among the positive mediators suggests a sequence-context dependence in the deletion-mediated pathway. However, this observational analysis does not establish the underlying biochemical mechanism, and further controlled synthesis experiments would be required to determine why C-ending contexts are preferentially represented. Figure~\ref{fig:contrib} compares the relative contributions of the significant pathways, and the estimated pathway coefficients with their confidence intervals are reported in Supplementary Figure S4. 
	
	\subsection{Engineering Interpretation}
	The pathway analysis provides information beyond the overall association and its direct and aggregate mediated components. By localizing the mediated effect to specific deletion-prone trinucleotide contexts, the analysis identifies sequence features that may warrant further consideration in sequence design and error-control strategies. The aggregate quantities above are signed signals, so positive and negative pathways may partially cancel; they should therefore be interpreted as net log-hazard components rather than as percentages of the total association. Such information could potentially be used to penalize high-risk contexts during sequence generation or to allocate additional redundancy to sequences with unfavorable damage profiles. These interpretations should, however, be viewed as design hypotheses rather than proven interventions, because the present analysis does not directly evaluate the effect of experimentally modifying these sequence contexts.
	
	\section{Conclusion}
	\label{sec:conclusion}
	This paper proposes a high-dimensional semiparametric mediation framework for identifying damage pathways linking sequence composition to DNA storage failure. By combining Cox regression, semiparametric mediator modeling, sparse selection, and joint testing, the method accommodates nonlinear covariate effects and identifies interpretable mediation signals. In a DNA aging experiment, several context-specific deletion pathways, especially in C-ending trinucleotide contexts, were associated with earlier storage-quality failure. These findings highlight structured damage mechanisms and provide candidates for sequence design and further experimental validation.
	
	\section{Acknowledgements}
	This work has been supported by National Key R\&D Program of China (2025YFC3409900), Tianjin Future Industry Science and Technology Major Project (25ZXWCSY00290).

	\bibliography{reference}

\newpage
\begin{center}
  Supplementary Materials for\\
  \large\bf “Identifying Damage Pathways Linking Sequence Composition to
Storage Failure in DNA Data Storage via High-Dimensional
Mediation Analysis”
\end{center}

\vspace*{0.2in}
	\appendix

\setcounter{section}{0}
\setcounter{table}{0}
\setcounter{figure}{0}
\renewcommand{\thesection}{S\arabic{section}}
\renewcommand{\thetable}{S\arabic{table}}
\renewcommand{\thefigure}{S\arabic{figure}}

	This Supplementary Material provides additional
	numerical analyses supporting the main manuscript. Section~\ref{sec:supp_identification} presents the
	counterfactual identification assumptions and methodological considerations
	underlying the proposed semiparametric mediation framework. Section~\ref{sec:simulation}
	reports additional simulation results under alternative covariate
	distributions and compares the proposed method with \textit{survHIMA} \citep{zhang2021}. Section~\ref{sec:supp_exposure_covariates}
	gives the formal definitions of the exposure variable, the sequence-derived
	covariates, and the high-dimensional damage mediators used in the DNA storage
	application. Section~\ref{sec:sensitivity}
	presents sensitivity analyses for the failure-threshold definitions used in
	the DNA storage application and examines the stability and sequence-context
	structure of the identified mediation pathways.

\section{Identification and Regularity Conditions}
	\label{sec:supp_identification}
	
	This section collects the causal identification assumptions and the
	technical regularity conditions underlying the proposed semiparametric
	mediation framework. These assumptions are stated separately from the
	main methodological development to keep the presentation in
	the main text focused on the statistical model, estimation procedure,
	and interpretation of the mediated pathways.
\subsection{Counterfactual Identification Assumptions}
	
	We first state the assumptions required for the counterfactual
	interpretation of the mediation pathways.
	
	The counterfactual interpretation  requires the standard assumptions
	of consistency, positivity, and sequential exchangeability. Specifically,
	we assume the following.
	
	\begin{enumerate}
		\item[(A1)] \textbf{Consistency.}
		If $X=x$ and $\bm M=\bm m$ are observed, then
		\[
		T=T(x,\bm m),
		\]
		and if $X=x$, then
		\[
		\bm M=\bm M(x).
		\]
		
		\item[(A2)] \textbf{No unmeasured exposure--outcome confounding.}
		For every $x$ and $\bm m$,
		\[
		T(x,\bm m)\perp X\mid\bm Z.
		\]
		
		\item[(A3)] \textbf{No unmeasured exposure--mediator confounding.}
		For every $x$,
		\[
		\bm M(x)\perp X\mid\bm Z.
		\]
		
		\item[(A4)] \textbf{No unmeasured mediator--outcome confounding.}
		For every $x$ and $\bm m$,
		\[
		T(x,\bm m)\perp \bm M
		\mid X=x,\bm Z.
		\]
		
		\item[(A5)] \textbf{No exposure-induced mediator--outcome confounding.}
		There is no post-exposure variable affected by $X$ that simultaneously
		confounds the relationship between $\bm M$ and $T$.
		
		\item[(A6)] \textbf{Positivity.}
		For all values of $\bm Z$ with positive probability, the exposure and
		mediator values involved in the counterfactual contrasts have positive
		conditional probability or density.
		
		\item[(A7)] \textbf{Structural invariance of the mediator equation.}
		The potential mediator obeys
		\[
		M_k(x)=\alpha_kx+g_k(\bm Z)+e_k,
		\]
		with the same exogenous error $e_k$ under different exposure interventions.
		Hence,
		\[
		M_k(x)-M_k(x^\ast)
		=
		\alpha_k(x-x^\ast).
		\]
		
		\item[(A8)] \textbf{No exposure--mediator interaction on the
			log-hazard scale.}
		The conditional counterfactual hazard has the form
		\[
		\lambda\{t\mid x,\bm m,\bm Z\}
		=
		\lambda_0(t)
		\exp\{
		\gamma x+\bm\beta^\top\bm m+g(\bm Z)
		\},
		\]
		so that the coefficient $\beta_k$ does not depend on $x$.
		
	\end{enumerate}
	\subsection{Deep Neural Network Function Class}
	\label{sec:supp_dnn_class}
	This section gives the formal definition of the deep neural network (DNN) function class used to approximate the nuisance functions $g$ and $g_k$ in the mediator and survival models.

	A DNN with architecture $(L, \mathbf{w})$ has $L+1$ layers, including an input layer, $L-1$ hidden layers and an output layer, and a width vector $\mathbf{w} = (w_1, w_2, \ldots, w_{L+1})$, where each integer $w_i$ denotes the number of neurons in the $i$-th layer. An $(L+1)$-layer DNN with width vector $\mathbf{w}$ is a composite function $g = g_L \circ g_{L-1} \circ \cdots \circ g_0 : \mathbb{R}^q \to \mathbb{R}$ recursively defined as
	\[
	h_0(z)=z,
	\]\[
	h_\ell(z)
	=
	\sigma_\ell(W_\ell h_{\ell-1}(z)+b_\ell),
	\quad
	\ell=1,\ldots,L,
	\]\[
	g(z)
	=
	W_{L+1}h_L(z)+b_{L+1},
	\]
	where the matrices \( W_\ell \in \mathbb{R}^{w_{\ell+1} \times w_\ell} \) and vectors \( b_\ell \in \mathbb{R}^{w_{\ell+1}} \), for $\ell = 1,\ldots,L+1$, denote the network parameters. The matrix entries \((W_\ell)_{i,j}\) represent the weights connecting the \( j \)-th neuron in layer \( \ell-1 \) to the \( i \)-th neuron in layer \( \ell \), and the bias vector entries \((b_\ell)_i\) are the corresponding bias terms. In our case, the input dimension of the DNN matches the covariate dimension, i.e., $h_0 \in \mathbb{R}^q$, while the output dimension is $w_{L+1} = 1$. The activation functions $\sigma(\cdot)$ are simple nonlinear transformations that operate componentwise, that is, for any vector $\mathbf{x} = (x_1, \dots, x_d)^{\top}$, we have $\sigma(\mathbf{x}) = (\sigma(x_1), \dots, \sigma(x_d))^{\top}$. Consequently, the intermediate mapping at the $l$-th layer can be viewed as a vector-valued function $\mathbf{g}_l = (g_{l1}, \dots, g_{l w_l})^{\top}: \mathbb{R}^{w_{l-1}} \to \mathbb{R}^{w_l}$ for $l = 2, \dots, L+1$. While various activation functions have been proposed in the deep learning literature, the rectified linear unit (ReLU) remains the most widely adopted \citep{glorot2011}. It is defined as: $\sigma(x) = \max\{x, 0\}.$ We explicitly restrict the hypothesis space to DNNs with uniformly bounded weights and biases, and further impose a sparsity constraint to prevent overfitting:
	\begin{equation*}
		\begin{aligned}
			G(L, \mathbf{w}) = &\{g : g \text{ is a DNN with } (L+1) \text{ layers and width vector } \mathbf{w} \text{ such that}\\
			&\max\{\|W_\ell\|_{\infty}, \|b_\ell\|_{\infty}\} \leq 1, \text{ for all } \ell = 1, \ldots, L+1\},
		\end{aligned}
	\end{equation*}
	where \(\|\cdot\|_{\infty}\) denotes the supremum norm of matrix or vector.
	In practice, the sizes of the learned parameter matrices $\mathbf{W}_\ell$ and vectors $\mathbf{b}_\ell$ rarely grow excessively large, provided that the parameters are initialized with relatively small values prior to stochastic gradient training. Without loss of generality, we restrict our theoretical analysis to DNNs whose matrices and vectors are uniformly bounded by one. Given that deep feedforward networks with fully connected layers involve a massive number of parameters, they tend to overfit. We therefore employ dropout regularization \citep{srivastava2014}, which randomly deactivates hidden neurons with a defined probability, and further restrict to a class of sparse neural networks indexed by sparsity level $s \in \mathbb{N}_+$ and uniform bound $D > 0$:
	\[G(L, s, \mathbf{w}, D) := \left\{ g \in G(L, \mathbf{w}) : \sum_{\ell=1}^L \left( \|W_\ell\|_0 + \|b_\ell\|_0\right)  \leq s, \|g\|_{\infty} \leq D \right\},\]
	where \(\|\cdot\|_0\) is the number of nonzero entries of matrix or vector.

	\subsection{Theoretical and Methodological Considerations}
	\label{sec:supp_regular_conditions}
	The proposed three-step procedure builds on established results for deep
	partially linear estimation and nonconcave penalized variable selection,
	whose theoretical properties have been studied under appropriate
	regularity conditions.
	The mediator-model estimation follows the DP2LM framework \citep{wang2024dp2lm},
	whereas estimation of the survival component is motivated by the deep
	partially linear Cox model \citep{zhong2022dplcox}. SCAD penalization is used in the
	second stage because of its favorable sparsity and reduced-bias properties
	for estimating nonzero coefficients \citep{sun2024dplc}. These existing theoretical
	results provide methodological support for the proposed procedure, while
	its finite-sample performance is evaluated through simulation studies in
	Section~\ref{sec:simulation}.   

	\section{Additional Simulation Results}
	\label{sec:simulation}
	This section provides additional simulation results to complement those reported in the main text. While the main text focuses on the representative setting in which the standardized covariate \(X\) and the covariates \(Z\) are generated from normal distributions, we further examine the performance of the proposed method under four alternative distributional settings: Laplace, Uniform, Normal--Uniform, and Uniform--Laplace. The remaining data-generating mechanisms are identical to those described in the main text. For each setting, we report mediator-specific selection frequencies, false discovery rate (FDR), statistical power, and bias and mean squared error (MSE) under different sample sizes and censoring rates. Results for \textit{survHIMA}~\citep{zhang2021} are included as a benchmark. These additional experiments assess the robustness of the proposed method to departures from the normal covariate distribution.
	
	\subsection{Detailed Results under Alternative Distributional Settings}
	Tables~\ref{tab:laplace-frequency}--\ref{tab:laplace-bias-mse}, \ref{tab:uniform-frequency}--\ref{tab:uniform-bias-mse}, \ref{tab:normal-uniform-frequency}--\ref{tab:normal-uniform-bias-mse}, and~\ref{tab:uniform-laplace-frequency}--\ref{tab:uniform-laplace-bias-mse} report the selection, FDR and power, and bias and MSE results for the four alternative distributional settings. Within each setting, the frequency table gives the number of times each truly active mediator was selected in 200 replications. The FDR and power table summarizes overall selection performance. In the bias and MSE tables, the first row for each pathway coefficient reports bias and the parenthesized row reports MSE.

	Across the four alternative distributional settings, the proposed method screened all ten active mediators with consistently high frequency. At a 20\% censoring rate, screening frequencies ranged from 192 to 200 out of 200 replications, corresponding to empirical power between 96\% and 100\%. Under the more challenging 40\% censoring scenario, frequencies ranged from 179 to 200 (89.5\% to 100\%). The largest reductions occurred for mediators \(M_3\) and \(M_7\), which have weaker effects. Notably, \textit{survHIMA} exhibited more pronounced degradation for these weaker mediators, particularly \(M_7\), where screening frequencies dropped to 171--186 under high censoring across distributions. When the sample size increased to \(n=1000\), both methods achieved near-perfect screening (199--200 out of 200) for all mediators except \(M_7\), which remained the most challenging signal to detect.

	The effect of censoring is likewise visible at the level of individual mediators. Raising the censoring rate from 20\% to 40\% lowered the selection frequencies of the weaker signals, most notably \(M_3\) and \(M_7\), at \(n=500\), while the stronger signals remained at or near 200 in every configuration. Fewer observed failures provide less information for estimating the survival coefficients \(\beta\), yet the screening step retained high power throughout.

	To make the distributional labels unambiguous, the first component names the distribution of the exposure \(X_i\), and the second component names the common distribution of the six adjustment covariates \(Z_{ij}\), where \(i=1,\ldots,n\) indexes subjects and \(j=1,\ldots,6\) indexes adjustment covariates. The four settings are
	\[
	\begin{aligned}
		\text{Laplace:}\quad
		&X_i\sim\operatorname{Laplace}(0,0.5),
		&&Z_{ij}\sim\operatorname{Laplace}(0,0.5);\\
		\text{Uniform:}\quad
		&X_i\sim U(-1,1),
		&&Z_{ij}\sim U(-1,1);\\
		\text{Normal--Uniform:}\quad
		&X_i\sim N(0,1),
		&&Z_{ij}\sim U(-1,1);\\
		\text{Uniform--Laplace:}\quad
		&X_i\sim U(-1,1),
		&&Z_{ij}\sim\operatorname{Laplace}(0,0.5).
	\end{aligned}
	\]
	Here, \(U(a,b)\) denotes a uniform distribution on \([a,b]\), \(N(0,1)\) denotes a normal distribution with mean zero and variance one, and the second argument of \(\operatorname{Laplace}(0,0.5)\) is its scale parameter. The draws of \(X_i\) and \(Z_{ij}\) are independent across subjects and across covariate components. These supplementary comparisons evaluate the method under the stated changes in both distributional shape. Across these settings, selection frequencies are generally close to 200, with the largest reductions concentrated among the weaker signals (especially \(M_3\) and \(M_7\)) at smaller sample size and heavier censoring.

	\subsection{Cross-Distribution Comparison of Detection Performance}
	The operating characteristics reported in Tables~\ref{tab:laplace-fdr-power}, \ref{tab:uniform-fdr-power}, \ref{tab:normal-uniform-fdr-power}, and~\ref{tab:uniform-laplace-fdr-power} permit a direct comparison across the four alternative settings. The proposed method yielded FDR values between 0 and 0.0014, whereas the FDR of \textit{survHIMA} ranged from 0.0348 to 0.0703. Both methods had high power: the proposed method ranged from 0.9825 to 1.0000 and \textit{survHIMA} ranged from 0.9780 to 0.9990.

	The proposed method's FDR control is substantially tighter than that of \textit{survHIMA} across all distributional settings. The maximum observed FDR for the proposed method was 0.0014 (less than 0.15\%), far below the nominal 5\% level, whereas \textit{survHIMA} exceeded 5\% in several configurations, reaching as high as 7.03\% under the Laplace distribution with $n=1000$ and 20\% censoring. Both methods maintained high power across all settings, but the proposed method demonstrated more consistent performance under the higher censoring level. At 40\% censoring with $n=500$, the proposed method's power remained above 98\% in all four distributions, whereas \textit{survHIMA} showed slightly greater variability, ranging from 97.8\% to 98.7\%. The power advantage of the proposed method is most evident for mediators with weaker effect sizes, such as $M_3$ and $M_7$. The proposed method achieves consistent bias and MSE reductions across all distributional settings. The average absolute bias reduction ranged from 68\% to 75\% across the four settings, with the largest improvements observed for mediators with larger effect sizes ($M_6$, $M_9$, $M_{10}$). The MSE reductions were somewhat more modest, averaging between 8\% and 12\%, reflecting the trade-off between reduced bias and the additional variance introduced by the DNN estimation. Overall, the proposed method achieved smaller absolute bias and competitive or lower MSE in most simulation configurations. The improvement was more pronounced in bias than in MSE, reflecting the bias--variance trade-off introduced by flexible nonlinear nuisance estimation.

	Together, these results demonstrate that the proposed method is robust to departures from normality in the covariate distribution and maintains its advantages in FDR control, detection power, and estimation accuracy across a range of distributional settings.
		\begin{table}[H]
			\centering
			\caption{Selection frequencies of the true active mediators under the Laplace-distribution setting}
			\label{tab:laplace-frequency}
			\begin{tabular}{cccccc}
				\toprule
				& & \multicolumn{2}{c}{CR=20\%} & \multicolumn{2}{c}{CR=40\%} \\
				\cmidrule(lr){3-4} \cmidrule(lr){5-6}
				&  &  Proposed & survHIMA &  Proposed & survHIMA \\
				\midrule
				\multirow{10}{*}{\( n = 500 \)} 
				& \( M_1 \) & 200 & 200 & 200 & 200 \\
				& \( M_2 \) & 200 & 195 & 200 & 197 \\
				& \( M_3 \) & 197 & 197 & 187 & 196 \\
				& \( M_4 \) & 200 & 198 & 200 & 199 \\
				& \( M_5 \) & 200 & 200 & 198 & 200 \\
				& \( M_6 \) & 200 & 200 & 200 & 200 \\
				& \( M_7 \) & 198 & 177 & 196 & 174 \\
				& \( M_8 \) & 199 & 200 & 199 & 200 \\
				& \( M_9 \) & 200 & 200 & 200 & 200 \\
				& \( M_{10} \) & 200 & 200 & 200 & 200 \\
				\cmidrule(lr){1-6}
				\multirow{10}{*}{\( n = 1000 \)} 
				& \( M_1 \) & 200 & 200 & 200 & 200 \\
				& \( M_2 \) & 199 & 200 & 199 & 200 \\
				& \( M_3 \) & 199 & 200 & 199 & 200 \\
				& \( M_4 \) & 200 & 200 & 199 & 200 \\
				& \( M_5 \) & 199 & 200 & 199 & 200 \\
				& \( M_6 \) & 200 & 200 & 200 & 200 \\
				& \( M_7 \) & 200 & 194 & 199 & 198 \\
				& \( M_8 \) & 199 & 200 & 199 & 200 \\
				& \( M_9 \) & 200 & 200 & 200 & 200 \\
				& \( M_{10} \) & 200 & 200 & 200 & 200 \\
				\bottomrule
			\end{tabular}
		\end{table}
		
		\begin{table}[H]
			\centering
			\caption{False discovery rate and statistical power (Laplace Distribution)}
			\label{tab:laplace-fdr-power}
			\small
			\setlength{\tabcolsep}{6pt}
			\begin{tabular}{llrrrr}
				\toprule
				Metric & Method
				& \multicolumn{2}{c}{CR = 20\%}
				& \multicolumn{2}{c}{CR = 40\%} \\
				\cmidrule(lr){3-4} \cmidrule(lr){5-6}
				& & $n=500$ & $n=1000$ & $n=500$ & $n=1000$ \\
				\midrule
				\multirow{2}{*}{FDR}
				& Proposed & 0.00\% & 0.05\% & 0.10\% & 0.00\% \\
				& survHIMA & 5.01\% & 7.03\% & 4.11\% & 6.33\% \\
				\midrule
				\multirow{2}{*}{Power}
				& Proposed & 99.70\% & 99.80\% & 99.00\% & 99.70\% \\
				& survHIMA & 98.35\% & 99.70\% & 98.30\% & 99.90\% \\
				\bottomrule
			\end{tabular}
		\end{table}
		
		\begin{table}[H]
			\centering
			\footnotesize
			\setlength{\tabcolsep}{3pt}
			\caption{Bias and MSE (in parentheses) of the estimated mediation effects under the Laplace-distribution setting}
			\label{tab:laplace-bias-mse}
			\begin{tabular}{ccccccccc}
				\toprule
				& \multicolumn{4}{c}{CR = 20\%} & \multicolumn{4}{c}{CR = 40\%} \\
				\cmidrule(lr){2-5} \cmidrule(lr){6-9}
				& \multicolumn{2}{c}{n = 500} & \multicolumn{2}{c}{n = 1000} 
				& \multicolumn{2}{c}{n = 500} & \multicolumn{2}{c}{n = 1000} \\
				\cmidrule(lr){2-3} \cmidrule(lr){4-5} 
				\cmidrule(lr){6-7} \cmidrule(lr){8-9}
				& Proposed & survHIMA & Proposed & survHIMA & Proposed & survHIMA & Proposed & survHIMA \\
				\midrule
				\multirow{2}{*}{$\alpha_1 \beta_1$} 
				& -0.0066 & -0.0518 & -0.0242 & -0.0442 & -0.0072 & -0.0474 & -0.0268 & -0.0389 \\
				& (0.0033) & (0.0052) & (0.0023) & (0.0033) & (0.0041) & (0.0049) & (0.0024) & (0.0029) \\
				\cmidrule(lr){1-9}
				\multirow{2}{*}{$\alpha_2 \beta_2$} 
				& -0.0016 & -0.0329 & -0.0192 & -0.0313 & -0.0045 & -0.0338 & -0.0207 & -0.0280 \\
				& (0.0020) & (0.0026) & (0.0016) & (0.0019) & (0.0027) & (0.0029) & (0.0016) & (0.0016) \\
				\cmidrule(lr){1-9}
				\multirow{2}{*}{$\alpha_3 \beta_3$} 
				& -0.0098 & -0.0302 & -0.0121 & -0.0183 & -0.0116 & -0.0287 & -0.0131 & -0.0173 \\
				& (0.0016) & (0.0019) & (0.0009) & (0.0009) & (0.0021) & (0.0019) & (0.0011) & (0.0009) \\
				\cmidrule(lr){1-9}
				\multirow{2}{*}{$\alpha_4 \beta_4$} 
				& -0.0016 & -0.0411 & -0.0245 & -0.0401 & -0.0036 & -0.0393 & -0.0271 & -0.0372 \\
				& (0.0033) & (0.0039) & (0.0021) & (0.0027) & (0.0035) & (0.0040) & (0.0027) & (0.0025) \\
				\cmidrule(lr){1-9}
				\multirow{2}{*}{$\alpha_5 \beta_5$} 
				& 0.0064 & -0.0253 & -0.0102 & -0.0229 & 0.0066 & -0.0222 & -0.0096 & -0.0192 \\
				& (0.0022) & (0.0020) & (0.0012) & (0.0013) & (0.0025) & (0.0021) & (0.0013) & (0.0013) \\
				\cmidrule(lr){1-9}
				\multirow{2}{*}{$\alpha_6 \beta_6$} 
				& 0.0056 & -0.0662 & -0.0248 & -0.0544 & 0.0035 & -0.0608 & -0.0248 & -0.0435 \\
				& (0.0062) & (0.0091) & (0.0032) & (0.0050) & (0.0066) & (0.0084) & (0.0034) & (0.0041) \\
				\cmidrule(lr){1-9}
				\multirow{2}{*}{$\alpha_7 \beta_7$} 
				& -0.0036 & -0.0247 & -0.0116 & -0.0218 & -0.0051 & -0.0250 & -0.0155 & -0.0203 \\
				& (0.0022) & (0.0020) & (0.0010) & (0.0012) & (0.0029) & (0.0024) & (0.0014) & (0.0012) \\
				\cmidrule(lr){1-9}
				\multirow{2}{*}{$\alpha_8 \beta_8$} 
				& 0.0004 & -0.0367 & -0.0126 & -0.0287 & -0.0041 & -0.0359 & -0.0142 & -0.0238 \\
				& (0.0026) & (0.0034) & (0.0016) & (0.0018) & (0.0028) & (0.0033) & (0.0017) & (0.0016) \\
				\cmidrule(lr){1-9}
				\multirow{2}{*}{$\alpha_9 \beta_9$} 
				& -0.0009 & -0.0661 & -0.0242 & -0.0560 & -0.0008 & -0.0530 & -0.0259 & -0.0458 \\
				& (0.0058) & (0.0082) & (0.0035) & (0.0058) & (0.0070) & (0.0074) & (0.0040) & (0.0048) \\
				\cmidrule(lr){1-9}
				\multirow{2}{*}{$\alpha_{10} \beta_{10}$} 
				& 0.0083 & -0.0476 & -0.0159 & -0.0440 & 0.0058 & -0.0448 & -0.0166 & -0.0368 \\
				& (0.0041) & (0.0049) & (0.0020) & (0.0036) & (0.0043) & (0.0046) & (0.0022) & (0.0032) \\
				\bottomrule
			\end{tabular}
		\end{table}

		\begin{table}[H]
			\centering
			\caption{Selection frequencies of the true active mediators under the uniform-distribution setting}
			\label{tab:uniform-frequency}
			\begin{tabular}{cccccc}
				\toprule
				& & \multicolumn{2}{c}{CR=20\%} & \multicolumn{2}{c}{CR=40\%} \\
				\cmidrule(lr){3-4} \cmidrule(lr){5-6}
				&  &  Proposed & survHIMA &  Proposed & survHIMA \\
				\midrule
				\multirow{10}{*}{\( n = 500 \)} 
				& \( M_1 \) & 200 & 199 & 200 & 200 \\
				& \( M_2 \) & 200 & 197 & 199 & 193 \\
				& \( M_3 \) & 194 & 196 & 184 & 195 \\
				& \( M_4 \) & 200 & 198 & 200 & 197 \\
				& \( M_5 \) & 200 & 199 & 198 & 200 \\
				& \( M_6 \) & 200 & 199 & 200 & 200 \\
				& \( M_7 \) & 198 & 178 & 196 & 171 \\
				& \( M_8 \) & 200 & 199 & 200 & 200 \\
				& \( M_9 \) & 200 & 199 & 200 & 200 \\
				& \( M_{10} \) & 200 & 199 & 200 & 200 \\
				\cmidrule(lr){1-6}
				\multirow{10}{*}{\( n = 1000 \)} 
				& \( M_1 \) & 200 & 200 & 200 & 200 \\
				& \( M_2 \) & 200 & 200 & 200 & 200 \\
				& \( M_3 \) & 200 & 200 & 200 & 200 \\
				& \( M_4 \) & 200 & 200 & 200 & 200 \\
				& \( M_5 \) & 200 & 200 & 200 & 200 \\
				& \( M_6 \) & 200 & 200 & 200 & 200 \\
				& \( M_7 \) & 200 & 188 & 200 & 193 \\
				& \( M_8 \) & 200 & 200 & 200 & 200 \\
				& \( M_9 \) & 200 & 200 & 200 & 200 \\
				& \( M_{10} \) & 200 & 200 & 200 & 200 \\
				\bottomrule
			\end{tabular}
		\end{table}
		
		\begin{table}[H]
			\centering
			\caption{False discovery rate and statistical power (Uniform Distribution)}
			\label{tab:uniform-fdr-power}
			\small
			\setlength{\tabcolsep}{6pt}
			\begin{tabular}{llrrrr}
				\toprule
				Metric & Method
				& \multicolumn{2}{c}{CR = 20\%}
				& \multicolumn{2}{c}{CR = 40\%} \\
				\cmidrule(lr){3-4} \cmidrule(lr){5-6}
				& & $n=500$ & $n=1000$ & $n=500$ & $n=1000$ \\
				\midrule
				\multirow{2}{*}{FDR}
				& Proposed & 0.05\% & 0.00\% & 0.10\% & 0.05\% \\
				& survHIMA & 4.67\% & 5.33\% & 4.68\% & 6.70\% \\
				\midrule
				\multirow{2}{*}{Power}
				& Proposed & 99.60\% & 100.00\% & 98.85\% & 100.00\% \\
				& survHIMA & 98.15\% & 99.40\% & 97.80\% & 99.65\% \\
				\bottomrule
			\end{tabular}
		\end{table}
		
		\begin{table}[H]
			\centering
			\footnotesize
			\setlength{\tabcolsep}{3pt}
			\caption{Bias and MSE (in parentheses) of the estimated mediation effects under the uniform-distribution setting}
			\label{tab:uniform-bias-mse}
			\begin{tabular}{ccccccccc}
				\toprule
				& \multicolumn{4}{c}{CR = 20\%} & \multicolumn{4}{c}{CR = 40\%} \\
				\cmidrule(lr){2-5} \cmidrule(lr){6-9}
				& \multicolumn{2}{c}{n = 500} & \multicolumn{2}{c}{n = 1000} 
				& \multicolumn{2}{c}{n = 500} & \multicolumn{2}{c}{n = 1000} \\
				\cmidrule(lr){2-3} \cmidrule(lr){4-5} 
				\cmidrule(lr){6-7} \cmidrule(lr){8-9}
				& Proposed & survHIMA & Proposed & survHIMA & Proposed & survHIMA & Proposed & survHIMA \\
				\midrule
				\multirow{2}{*}{$\alpha_1 \beta_1$} 
				& 0.0047 & -0.0371 & -0.0109 & -0.0262 & 0.0044 & -0.0386 & -0.0148 & -0.0245 \\
				& (0.0041) & (0.0043) & (0.0021) & (0.0024) & (0.0051) & (0.0050) & (0.0024) & (0.0024) \\
				\cmidrule(lr){1-9}
				\multirow{2}{*}{$\alpha_2 \beta_2$} 
				& 0.0014 & -0.0241 & -0.0124 & -0.0221 & -0.0039 & -0.0251 & -0.0154 & -0.0209 \\
				& (0.0025) & (0.0024) & (0.0013) & (0.0014) & (0.0033) & (0.0026) & (0.0014) & (0.0014) \\
				\cmidrule(lr){1-9}
				\multirow{2}{*}{$\alpha_3 \beta_3$} 
				& -0.0042 & -0.0214 & -0.0077 & -0.0126 & -0.0105 & -0.0250 & -0.0101 & -0.0121 \\
				& (0.0017) & (0.0016) & (0.0008) & (0.0008) & (0.0020) & (0.0020) & (0.0010) & (0.0009) \\
				\cmidrule(lr){1-9}
				\multirow{2}{*}{$\alpha_4 \beta_4$} 
				& 0.0045 & -0.0304 & -0.0157 & -0.0285 & -0.0003 & -0.0317 & -0.0201 & -0.0279 \\
				& (0.0041) & (0.0035) & (0.0018) & (0.0021) & (0.0046) & (0.0037) & (0.0024) & (0.0022) \\
				\cmidrule(lr){1-9}
				\multirow{2}{*}{$\alpha_5 \beta_5$} 
				& 0.0134 & -0.0159 & -0.0022 & -0.0146 & 0.0124 & -0.0152 & -0.0031 & -0.0126 \\
				& (0.0025) & (0.0019) & (0.0012) & (0.0012) & (0.0026) & (0.0020) & (0.0014) & (0.0014) \\
				\cmidrule(lr){1-9}
				\multirow{2}{*}{$\alpha_6 \beta_6$} 
				& 0.0196 & -0.0433 & -0.0091 & -0.0358 & 0.0162 & -0.0452 & -0.0125 & -0.0288 \\
				& (0.0071) & (0.0077) & (0.0030) & (0.0039) & (0.0084) & (0.0081) & (0.0033) & (0.0031) \\
				\cmidrule(lr){1-9}
				\multirow{2}{*}{$\alpha_7 \beta_7$} 
				& -0.0026 & -0.0180 & -0.0095 & -0.0144 & -0.0031 & -0.0152 & -0.0126 & -0.0147 \\
				& (0.0026) & (0.0021) & (0.0010) & (0.0010) & (0.0032) & (0.0024) & (0.0012) & (0.0011) \\
				\cmidrule(lr){1-9}
				\multirow{2}{*}{$\alpha_8 \beta_8$} 
				& 0.0135 & -0.0238 & -0.0057 & -0.0219 & 0.0126 & -0.0246 & -0.0086 & -0.0195 \\
				& (0.0036) & (0.0032) & (0.0013) & (0.0019) & (0.0043) & (0.0035) & (0.0015) & (0.0017) \\
				\cmidrule(lr){1-9}
				\multirow{2}{*}{$\alpha_9 \beta_9$} 
				& 0.0217 & -0.0441 & -0.0057 & -0.0356 & 0.0190 & -0.0407 & -0.0114 & -0.0290 \\
				& (0.0081) & (0.0070) & (0.0033) & (0.0042) & (0.0084) & (0.0075) & (0.0041) & (0.0041) \\
				\cmidrule(lr){1-9}
				\multirow{2}{*}{$\alpha_{10} \beta_{10}$} 
				& 0.0209 & -0.0340 & -0.0009 & -0.0279 & 0.0164 & -0.0344 & -0.0032 & -0.0252 \\
				& (0.0055) & (0.0042) & (0.0022) & (0.0028) & (0.0057) & (0.0045) & (0.0023) & (0.0029) \\
				\bottomrule
			\end{tabular}
		\end{table}

		\begin{table}[H]
			\centering
			\caption{Selection frequencies of the true active mediators under the normal--uniform-distribution setting}
			\label{tab:normal-uniform-frequency}
			\begin{tabular}{cccccc}
				\toprule
				& & \multicolumn{2}{c}{CR=20\%} & \multicolumn{2}{c}{CR=40\%} \\
				\cmidrule(lr){3-4} \cmidrule(lr){5-6}
				&  &  Proposed & survHIMA &  Proposed & survHIMA \\
				\midrule
				\multirow{10}{*}{\( n = 500 \)} 
				& \( M_1 \) & 200 & 200 & 200 & 199 \\
				& \( M_2 \) & 200 & 198 & 200 & 197 \\
				& \( M_3 \) & 198 & 197 & 194 & 191 \\
				& \( M_4 \) & 200 & 199 & 200 & 199 \\
				& \( M_5 \) & 200 & 200 & 199 & 199 \\
				& \( M_6 \) & 200 & 200 & 200 & 199 \\
				& \( M_7 \) & 200 & 190 & 197 & 183 \\
				& \( M_8 \) & 200 & 200 & 200 & 199 \\
				& \( M_9 \) & 200 & 200 & 200 & 199 \\
				& \( M_{10} \) & 200 & 200 & 200 & 199 \\
				\cmidrule(lr){1-6}
				\multirow{10}{*}{\( n = 1000 \)} 
				& \( M_1 \) & 200 & 200 & 200 & 200 \\
				& \( M_2 \) & 200 & 200 & 200 & 200 \\
				& \( M_3 \) & 200 & 200 & 200 & 200 \\
				& \( M_4 \) & 200 & 200 & 200 & 200 \\
				& \( M_5 \) & 200 & 200 & 200 & 200 \\
				& \( M_6 \) & 200 & 200 & 200 & 200 \\
				& \( M_7 \) & 200 & 195 & 200 & 196 \\
				& \( M_8 \) & 200 & 200 & 200 & 200 \\
				& \( M_9 \) & 200 & 200 & 200 & 200 \\
				& \( M_{10} \) & 200 & 200 & 200 & 200 \\
				\bottomrule
			\end{tabular}
		\end{table}
		
		\begin{table}[H]
			\centering
			\caption{False discovery rate and statistical power (Normal-Uniform Distribution)}
			\label{tab:normal-uniform-fdr-power}
			\small
			\setlength{\tabcolsep}{6pt}
			\begin{tabular}{llrrrr}
				\toprule
				Metric & Method
				& \multicolumn{2}{c}{CR = 20\%}
				& \multicolumn{2}{c}{CR = 40\%} \\
				\cmidrule(lr){3-4} \cmidrule(lr){5-6}
				& & $n=500$ & $n=1000$ & $n=500$ & $n=1000$ \\
				\midrule
				\multirow{2}{*}{FDR}
				& Proposed & 0.05\% & 0.09\% & 0.14\% & 0.05\% \\
				& survHIMA & 4.08\% & 5.74\% & 3.48\% & 5.84\% \\
				\midrule
				\multirow{2}{*}{Power}
				& Proposed & 99.90\% & 100.00\% & 99.50\% & 100.00\% \\
				& survHIMA & 99.20\% & 99.75\% & 98.20\% & 99.80\% \\
				\bottomrule
			\end{tabular}
		\end{table}
		
		\begin{table}[H]
			\centering
			\footnotesize
			\setlength{\tabcolsep}{3pt}
			\caption{Bias and MSE (in parentheses) of the estimated mediation effects under the normal--uniform-distribution setting}
			\label{tab:normal-uniform-bias-mse}
			\begin{tabular}{ccccccccc}
				\toprule
				& \multicolumn{4}{c}{CR = 20\%} & \multicolumn{4}{c}{CR = 40\%} \\
				\cmidrule(lr){2-5} \cmidrule(lr){6-9}
				& \multicolumn{2}{c}{n = 500} & \multicolumn{2}{c}{n = 1000} 
				& \multicolumn{2}{c}{n = 500} & \multicolumn{2}{c}{n = 1000} \\
				\cmidrule(lr){2-3} \cmidrule(lr){4-5} 
				\cmidrule(lr){6-7} \cmidrule(lr){8-9}
				& Proposed & survHIMA & Proposed & survHIMA & Proposed & survHIMA & Proposed & survHIMA \\
				\midrule
				\multirow{2}{*}{$\alpha_1 \beta_1$} 
				& 0.0056 & -0.0437 & -0.0103 & -0.0328 & 0.0011 & -0.0456 & -0.0134 & -0.0325 \\
				& (0.0033) & (0.0039) & (0.0014) & (0.0021) & (0.0035) & (0.0043) & (0.0018) & (0.0023) \\
				\cmidrule(lr){1-9}
				\multirow{2}{*}{$\alpha_2 \beta_2$} 
				& 0.0099 & -0.0300 & -0.0040 & -0.0221 & 0.0074 & -0.0354 & -0.0068 & -0.0228 \\
				& (0.0024) & (0.0023) & (0.0011) & (0.0013) & (0.0024) & (0.0028) & (0.0013) & (0.0014) \\
				\cmidrule(lr){1-9}
				\multirow{2}{*}{$\alpha_3 \beta_3$} 
				& 0.0069 & -0.0192 & -0.0066 & -0.0186 & 0.0049 & -0.0215 & -0.0092 & -0.0203 \\
				& (0.0014) & (0.0011) & (0.0006) & (0.0007) & (0.0017) & (0.0014) & (0.0007) & (0.0008) \\
				\cmidrule(lr){1-9}
				\multirow{2}{*}{$\alpha_4 \beta_4$} 
				& 0.0176 & -0.0310 & -0.0056 & -0.0271 & 0.0125 & -0.0380 & -0.0092 & -0.0293 \\
				& (0.0031) & (0.0027) & (0.0010) & (0.0015) & (0.0035) & (0.0033) & (0.0012) & (0.0017) \\
				\cmidrule(lr){1-9}
				\multirow{2}{*}{$\alpha_5 \beta_5$} 
				& 0.0089 & -0.0233 & -0.0042 & -0.0185 & 0.0071 & -0.0236 & -0.0059 & -0.0197 \\
				& (0.0021) & (0.0017) & (0.0007) & (0.0008) & (0.0027) & (0.0020) & (0.0008) & (0.0009) \\
				\cmidrule(lr){1-9}
				\multirow{2}{*}{$\alpha_6 \beta_6$} 
				& 0.0241 & -0.0444 & -0.0097 & -0.0410 & 0.0218 & -0.0483 & -0.0148 & -0.0409 \\
				& (0.0063) & (0.0051) & (0.0020) & (0.0031) & (0.0070) & (0.0059) & (0.0024) & (0.0032) \\
				\cmidrule(lr){1-9}
				\multirow{2}{*}{$\alpha_7 \beta_7$} 
				& 0.0054 & -0.0236 & -0.0029 & -0.0167 & 0.0025 & -0.0262 & -0.0057 & -0.0167 \\
				& (0.0018) & (0.0016) & (0.0008) & (0.0008) & (0.0020) & (0.0020) & (0.0009) & (0.0010) \\
				\cmidrule(lr){1-9}
				\multirow{2}{*}{$\alpha_8 \beta_8$} 
				& 0.0176 & -0.0231 & -0.0031 & -0.0220 & 0.0171 & -0.0241 & -0.0049 & -0.0214 \\
				& (0.0030) & (0.0019) & (0.0008) & (0.0012) & (0.0034) & (0.0024) & (0.0010) & (0.0013) \\
				\cmidrule(lr){1-9}
				\multirow{2}{*}{$\alpha_9 \beta_9$} 
				& 0.0388 & -0.0337 & 0.0010 & -0.0332 & 0.0410 & -0.0314 & -0.0043 & -0.0309 \\
				& (0.0075) & (0.0047) & (0.0024) & (0.0027) & (0.0087) & (0.0050) & (0.0028) & (0.0029) \\
				\cmidrule(lr){1-9}
				\multirow{2}{*}{$\alpha_{10} \beta_{10}$} 
				& 0.0285 & -0.0301 & -0.0035 & -0.0328 & 0.0251 & -0.0313 & -0.0056 & -0.0306 \\
				& (0.0042) & (0.0029) & (0.0013) & (0.0021) & (0.0042) & (0.0029) & (0.0015) & (0.0020) \\
				\bottomrule
			\end{tabular}
		\end{table}

		\begin{table}[H]
			\centering
			\caption{Selection frequencies of the true active mediators under the uniform--Laplace-distribution setting}
			\label{tab:uniform-laplace-frequency}
			\begin{tabular}{cccccc}
				\toprule
				& & \multicolumn{2}{c}{CR=20\%} & \multicolumn{2}{c}{CR=40\%} \\
				\cmidrule(lr){3-4} \cmidrule(lr){5-6}
				&  &  Proposed & survHIMA &  Proposed & survHIMA \\
				\midrule
				\multirow{10}{*}{\( n = 500 \)} 
				& \( M_1 \) & 200 & 200 & 200 & 200 \\
				& \( M_2 \) & 200 & 196 & 197 & 196 \\
				& \( M_3 \) & 192 & 197 & 179 & 194 \\
				& \( M_4 \) & 200 & 199 & 199 & 198 \\
				& \( M_5 \) & 200 & 200 & 197 & 200 \\
				& \( M_6 \) & 200 & 200 & 200 & 200 \\
				& \( M_7 \) & 197 & 171 & 194 & 176 \\
				& \( M_8 \) & 200 & 200 & 199 & 200 \\
				& \( M_9 \) & 200 & 200 & 200 & 200 \\
				& \( M_{10} \) & 200 & 200 & 200 & 200 \\
				\cmidrule(lr){1-6}
				\multirow{10}{*}{\( n = 1000 \)} 
				& \( M_1 \) & 200 & 200 & 200 & 200 \\
				& \( M_2 \) & 199 & 200 & 199 & 200 \\
				& \( M_3 \) & 199 & 199 & 199 & 200 \\
				& \( M_4 \) & 200 & 200 & 199 & 200 \\
				& \( M_5 \) & 199 & 200 & 199 & 200 \\
				& \( M_6 \) & 200 & 200 & 200 & 200 \\
				& \( M_7 \) & 199 & 189 & 199 & 192 \\
				& \( M_8 \) & 199 & 200 & 199 & 200 \\
				& \( M_9 \) & 200 & 200 & 200 & 200 \\
				& \( M_{10} \) & 200 & 200 & 200 & 200 \\
				\bottomrule
			\end{tabular}
		\end{table}
		
		\begin{table}[H]
			\centering
			\caption{False discovery rate and statistical power (Uniform-Laplace Distribution)}
			\label{tab:uniform-laplace-fdr-power}
			\small
			\setlength{\tabcolsep}{6pt}
			\begin{tabular}{llrrrr}
				\toprule
				Metric & Method
				& \multicolumn{2}{c}{CR = 20\%}
				& \multicolumn{2}{c}{CR = 40\%} \\
				\cmidrule(lr){3-4} \cmidrule(lr){5-6}
				& & $n=500$ & $n=1000$ & $n=500$ & $n=1000$ \\
				\midrule
				\multirow{2}{*}{FDR}
				& Proposed & 0.00\% & 0.09\% & 0.05\% & 0.00\% \\
				& survHIMA & 4.27\% & 6.39\% & 4.31\% & 6.50\% \\
				\midrule
				\multirow{2}{*}{Power}
				& Proposed & 99.45\% & 99.75\% & 98.25\% & 99.70\% \\
				& survHIMA & 98.15\% & 99.40\% & 98.20\% & 99.60\% \\
				\bottomrule
			\end{tabular}
		\end{table}
		
		\begin{table}[H]
			\centering
			\footnotesize
			\setlength{\tabcolsep}{3pt}
			\caption{Bias and MSE (in parentheses) of the estimated mediation effects under the uniform--Laplace-distribution setting}
			\label{tab:uniform-laplace-bias-mse}
			\begin{tabular}{ccccccccc}
				\toprule
				& \multicolumn{4}{c}{CR = 20\%} & \multicolumn{4}{c}{CR = 40\%} \\
				\cmidrule(lr){2-5} \cmidrule(lr){6-9}
				& \multicolumn{2}{c}{n = 500} & \multicolumn{2}{c}{n = 1000} 
				& \multicolumn{2}{c}{n = 500} & \multicolumn{2}{c}{n = 1000} \\
				\cmidrule(lr){2-3} \cmidrule(lr){4-5} 
				\cmidrule(lr){6-7} \cmidrule(lr){8-9}
				& Proposed & survHIMA & Proposed & survHIMA & Proposed & survHIMA & Proposed & survHIMA \\
				\midrule
				\multirow{2}{*}{$\alpha_1 \beta_1$} 
				& -0.0103 & -0.0510 & -0.0255 & -0.0416 & -0.0090 & -0.0473 & -0.0274 & -0.0361 \\
				& (0.0042) & (0.0054) & (0.0026) & (0.0035) & (0.0051) & (0.0054) & (0.0028) & (0.0031) \\
				\cmidrule(lr){1-9}
				\multirow{2}{*}{$\alpha_2 \beta_2$} 
				& -0.0063 & -0.0321 & -0.0233 & -0.0313 & -0.0124 & -0.0324 & -0.0252 & -0.0282 \\
				& (0.0027) & (0.0029) & (0.0019) & (0.0020) & (0.0037) & (0.0030) & (0.0020) & (0.0018) \\
				\cmidrule(lr){1-9}
				\multirow{2}{*}{$\alpha_3 \beta_3$} 
				& -0.0113 & -0.0286 & -0.0148 & -0.0179 & -0.0160 & -0.0278 & -0.0166 & -0.0172 \\
				& (0.0019) & (0.0020) & (0.0011) & (0.0010) & (0.0025) & (0.0021) & (0.0013) & (0.0010) \\
				\cmidrule(lr){1-9}
				\multirow{2}{*}{$\alpha_4 \beta_4$} 
				& -0.0059 & -0.0414 & -0.0276 & -0.0381 & -0.0108 & -0.0374 & -0.0297 & -0.0359 \\
				& (0.0038) & (0.0043) & (0.0025) & (0.0028) & (0.0043) & (0.0043) & (0.0031) & (0.0027) \\
				\cmidrule(lr){1-9}
				\multirow{2}{*}{$\alpha_5 \beta_5$} 
				& 0.0047 & -0.0256 & -0.0116 & -0.0229 & 0.0040 & -0.0210 & -0.0105 & -0.0193 \\
				& (0.0022) & (0.0022) & (0.0015) & (0.0016) & (0.0025) & (0.0023) & (0.0016) & (0.0015) \\
				\cmidrule(lr){1-9}
				\multirow{2}{*}{$\alpha_6 \beta_6$} 
				& 0.0011 & -0.0642 & -0.0291 & -0.0546 & -0.0001 & -0.0580 & -0.0290 & -0.0452 \\
				& (0.0066) & (0.0099) & (0.0039) & (0.0058) & (0.0083) & (0.0091) & (0.0039) & (0.0046) \\
				\cmidrule(lr){1-9}
				\multirow{2}{*}{$\alpha_7 \beta_7$} 
				& -0.0094 & -0.0237 & -0.0165 & -0.0221 & -0.0126 & -0.0233 & -0.0194 & -0.0209 \\
				& (0.0027) & (0.0024) & (0.0015) & (0.0013) & (0.0036) & (0.0028) & (0.0015) & (0.0013) \\
				\cmidrule(lr){1-9}
				\multirow{2}{*}{$\alpha_8 \beta_8$} 
				& 0.0031 & -0.0370 & -0.0153 & -0.0311 & -0.0001 & -0.0332 & -0.0164 & -0.0263 \\
				& (0.0033) & (0.0041) & (0.0018) & (0.0024) & (0.0040) & (0.0040) & (0.0020) & (0.0020) \\
				\cmidrule(lr){1-9}
				\multirow{2}{*}{$\alpha_9 \beta_9$} 
				& -0.0053 & -0.0667 & -0.0272 & -0.0568 & -0.0032 & -0.0546 & -0.0284 & -0.0460 \\
				& (0.0068) & (0.0090) & (0.0042) & (0.0064) & (0.0085) & (0.0086) & (0.0047) & (0.0053) \\
				\cmidrule(lr){1-9}
				\multirow{2}{*}{$\alpha_{10} \beta_{10}$} 
				& 0.0068 & -0.0477 & -0.0155 & -0.0433 & 0.0004 & -0.0447 & -0.0165 & -0.0371 \\
				& (0.0052) & (0.0056) & (0.0024) & (0.0039) & (0.0052) & (0.0055) & (0.0025) & (0.0036) \\
				\bottomrule
			\end{tabular}
		\end{table}
	
	\section{Definition of Exposure and Covariates}	
	\label{sec:supp_exposure_covariates}
	
\begin{figure}[H]
		\centering
		\includegraphics[width=1\textwidth]{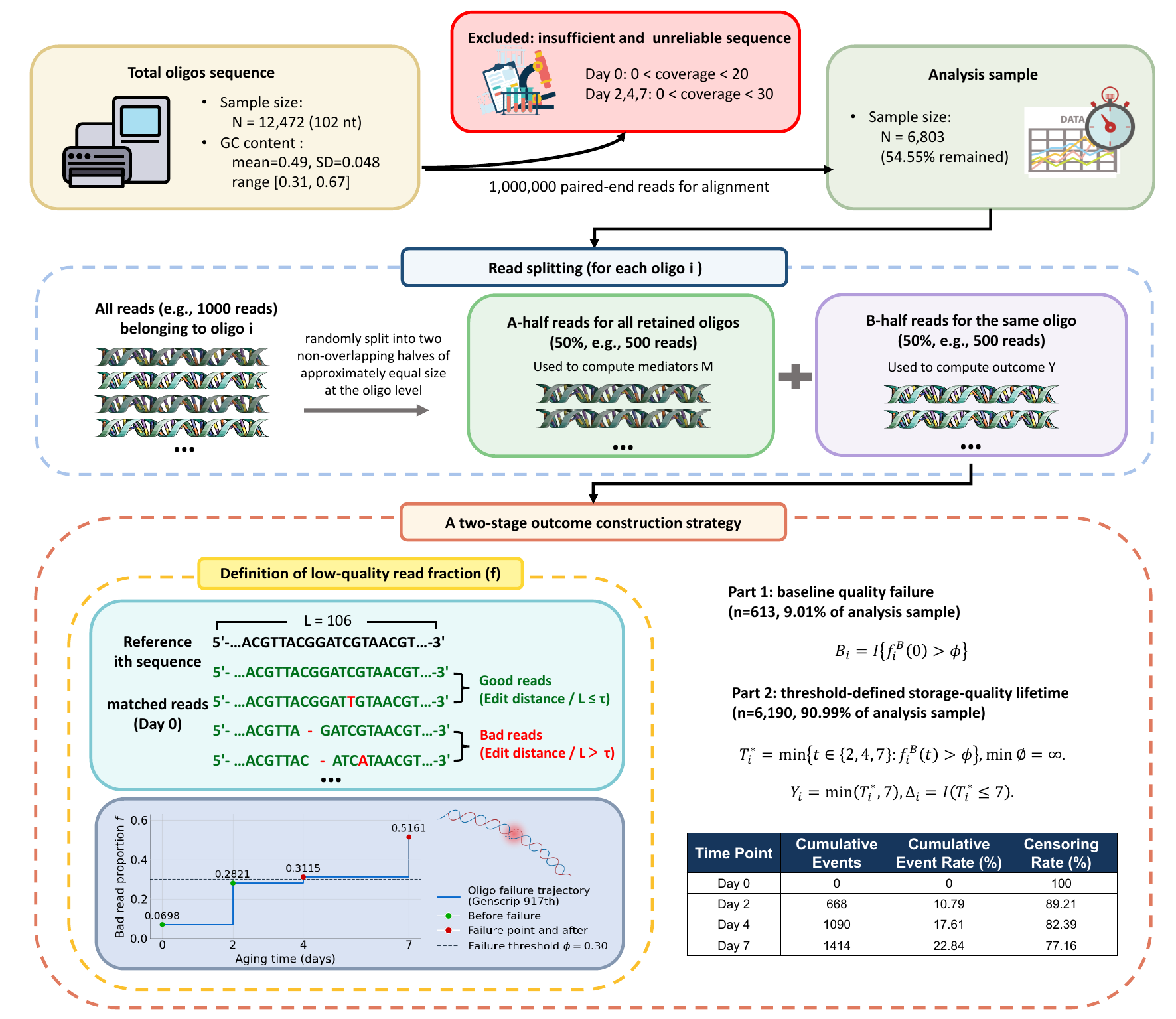}
		\caption{Data extract workflow. The illustration of DNA reads was modified from Chilton (2026) \citep{chilton2026dna}, and the illustration of damaged DNA was modified from Khdhr \citep{khdhr2026damaged}.}
		\label{fig:data_extract}
	\end{figure}

	This section provides formal definitions of the exposure variable $X$ and the covariate vector $\mathbf{Z}$ used in the proposed mediation framework. All quantities are derived from the design sequences and baseline sequencing data described in Section~4 of the main text. Figure~\ref{fig:data_extract} summarizes the three-stage extraction workflow that produces the exposure, mediators, covariates, and outcome from the design oligonucleotide pool.

	\subsection{Exposure: GC Content}
	
	For oligonucleotide $i$ with design sequence $s_i = (s_{i,0}, s_{i,1}, \ldots, s_{i,L_i-1})$ of length $L_i$, the GC content is defined as
	\[
	X_i = \frac{1}{L_i} \sum_{j=0}^{L_i-1} \mathbf{1}\{s_{i,j} \in \{G, C\}\},
	\]
	where $\mathbf{1}\{\cdot\}$ denotes the indicator function. Thus, $X_i \in [0,1]$ represents the proportion of guanine and cytosine bases in the design sequence. In the analysis, $X_i$ is centered and scaled to have mean zero and unit variance across the sample.
	
	The choice of GC content as the primary exposure is motivated by established findings in DNA synthesis and storage. High GC content is known to affect synthesis efficiency through secondary structure formation and to influence coverage bias in PCR amplification \citep{aird2011,press2020}. The sequences in the Genscript pool exhibit substantial GC variation (mean 0.493, standard deviation 0.048), providing adequate exposure contrast for mediation analysis.
	
	\subsection{Covariates}
	
	The covariate vector $\mathbf{Z}_i = (Z_{i,1}, Z_{i,2}, Z_{i,3}, Z_{i,4})^\top \in \mathbb{R}^4$ comprises four sequence-derived characteristics that adjust for potential confounding. All covariates are computed from the design sequence and remain constant throughout the aging experiment.
	
	\paragraph{$Z_1$: Maximum Homopolymer Length (maxHP)}
	A homopolymer run is a maximal contiguous subsequence of identical nucleotides. Maximum homopolymer length is defined as
	\[
	Z_{i,1} = \text{maxHP}(s_i) = \max_{0 \le j < L_i} \ell(j),
	\]
	where $\ell(j)$ denotes the length of the homopolymer run containing position $j$. Homopolymer runs are problematic in DNA synthesis and sequencing because polymerase slippage can lead to insertion or deletion errors \citep{IHGSC2001}. The maximum homopolymer length is a standard constraint parameter in DNA storage sequence design and is therefore included as a covariate to control for synthesis-related confounding.
	
	\paragraph{$Z_2$: Homopolymer Load (hp\_load)}
	
	Homopolymer load counts the number of homopolymer runs of length at least 3:
	\[
	Z_{i,2} = \text{hp\_load}(s_i) = \#\{\text{maximal homopolymer runs of length} \ge 3 \text{ in } s_i\}.
	\]
	This covariate quantifies the overall burden of homopolymer structure in the sequence. Whereas $Z_1$ captures the length of the longest run, $Z_2$ measures how frequently problematic homopolymer structures occur throughout the sequence. Homopolymer runs are known to cause elevated error rates in both DNA synthesis and sequencing due to polymerase slippage \citep{IHGSC2001,quince2011}.
	
	\paragraph{$Z_3$: 3-mer Shannon Entropy (ent3)}
	
	The 3-mer Shannon entropy~\citep{shannon1948} quantifies the compositional complexity of the sequence at the trinucleotide level. Let $\mathcal{T} = \{AAA, AAC, \ldots, TTT\}$ denote the set of all 64 possible trinucleotides. For each $t \in \mathcal{T}$, let $n_i(t)$ denote the number of occurrences of trinucleotide $t$ in sequence $s_i$. The total number of trinucleotides in the sequence is $N_i = L_i - 2$ (for a sequence of length $L_i$, there are $L_i - 2$ overlapping trinucleotide positions). The empirical frequency of trinucleotide $t$ is
	\[
	p_i(t) = \frac{n_i(t)}{N_i}.
	\]
	The 3-mer Shannon entropy is then defined as
	\[
	Z_{i,3} = \text{ent3}(s_i) = -\sum_{t \in \mathcal{T}} p_i(t) \log_2 p_i(t),
	\]
	where by convention $0 \log_2 0 = 0$. The entropy ranges from 0 (all trinucleotides are identical) to $\log_2 64 = 6$ bits (all trinucleotides occur with equal frequency). Lower entropy indicates a less diverse trinucleotide composition, which may be associated with reduced synthesis fidelity or increased susceptibility to context-dependent damage. The use of trinucleotide entropy to quantify sequence complexity is well established in DNA sequence analysis \citep{schmitt1997}.
	
	\paragraph{$Z_4$: GC Heterogeneity (GC\_het)}
	GC heterogeneity quantifies local variation in GC content using a sliding-window approach. For a sequence $s_i$ of length $L_i$, we partition the sequence into non-overlapping windows of width $w = 12$ nucleotides, discarding any remainder that does not form a complete window. Let $K_i = \lfloor L_i / w \rfloor$ denote the number of complete windows. For the $k$-th window ($k = 0, 1, \ldots, K_i - 1$), which spans positions $[kw, (k+1)w)$, the local GC proportion is
	\[
	g_{i,k} = \frac{1}{w} \sum_{j=kw}^{(k+1)w-1} \mathbf{1}\{s_{i,j} \in \{G, C\}\}.
	\]
	The mean local GC proportion is $\bar{g}_i = K_i^{-1} \sum_{k=0}^{K_i-1} g_{i,k}$. GC heterogeneity is then defined as the standard deviation of the local GC proportions:
	\[
	Z_{i,4} = \text{GC\_het}(s_i) = \sqrt{\frac{1}{K_i} \sum_{k=0}^{K_i-1} (g_{i,k} - \bar{g}_i)^2}.
	\]
	When the sequence length is an exact multiple of $w$, $\bar{g}_i$ equals the global GC content $X_i$. The window width $w = 12$ is chosen to match the scale at which sequence constraints are typically enforced in DNA storage coding schemes \citep{press2020}. GC heterogeneity is included as a covariate because localized clusters of G and C bases can act as hotspots for oxidative damage \citep{burrows1998}.

	\begin{table}[H]
		\centering
		\footnotesize
		\setlength{\tabcolsep}{3pt}
		\caption{Summary statistics of sequence covariates}
		\label{tab:covariate_summary}
		\begin{tabular}{lccc}
			\toprule
			Covariate & Mean & Std. Dev. & Range \\
			\midrule
			Maximum homopolymer length (maxHP)  & 3.99 & 0.94 & [2.00, 10.00] \\
			Homopolymer load (hp\_load)         & 4.46 & 1.88 & [0.00, 14.00] \\
			Trinucleotide entropy (ent3)        & 5.47 & 0.10 & [5.01, 5.78] \\
			GC heterogeneity (GC\_het)          & 0.13 & 0.03 & [0.03, 0.28] \\
			\bottomrule
		\end{tabular}
	\end{table}
	Table~\ref{tab:covariate_summary} presents summary statistics for the four sequence-derived covariates used in the mediation analysis. Maximum homopolymer length averaged 3.99 (range 2--10), homopolymer load averaged 4.46 runs per sequence (range 0--14), trinucleotide entropy averaged 5.47 bits (range 5.01--5.78, near the theoretical maximum of 6 bits), and GC heterogeneity averaged 0.13 (range 0.03--0.28).

	\section{Storage-Quality Failure Threshold Sensitivity}
	\label{sec:sensitivity}
	We calibrated the read-level threshold $\tau$ and oligo-level threshold $\phi$ using a grid search. We evaluated combinations formed by
	\[
	\begin{aligned}
	\tau &\in \{0.01, 0.03, 0.05, 0.07, 0.10\},\\
	\phi &\in \{0.02, 0.025, 0.03, 0.05, 0.10, 0.15, 0.20, 0.25, 0.30, 0.40, 0.50\}.
	\end{aligned}
	\]
		
	For each candidate threshold pair $(\tau,\phi)$, a read is first classified as
	low quality when its normalized Levenshtein edit distance exceeds $\tau$.
	For each oligo, the resulting fraction of low-quality reads at day 0 is then
	compared with $\phi$. Oligos for which this fraction exceeds $\phi$ are
	classified as baseline quality failures. This is an operational classification
	based on observed day-0 read quality and should not be interpreted as direct
	evidence of failure during chemical DNA synthesis, because the measured read
	quality can also reflect downstream amplification, sequencing, and alignment.
	These baseline quality failures are excluded before construction of the
	storage-lifetime outcome rather than being treated as right-censored
	observations. Therefore,
	the sample size reported in the table is threshold-specific,
	\[
	n(\tau,\phi)=\#\{\text{oligos retained under }(\tau,\phi)\},
	\]
	and is not the fixed size of the original sequencing pool. Storage-quality
	events and censoring rates are calculated within this retained analysis set.
	Increasing $\phi$ makes the baseline quality-failure criterion harder to
	satisfy, so fewer oligos are excluded and $n$ generally increases. Increasing
	$\tau$ also reduces the number of reads classified as low quality and can
	therefore increase the number of oligos retained.
	
	Table~\ref{tab:tau-sensitivity} presents the admissible $\phi$ values and corresponding event counts across different $\tau$ levels. As $\tau$ increases, a read must have a larger normalized edit distance to be classified as low quality. Consequently, fewer reads are labeled low quality at a fixed $\phi$, and lower $\phi$ values are needed to retain an adequate number of storage-quality failure events. The choice of the $\tau$ grid $\{0.01, 0.03, 0.05, 0.07, 0.10\}$ follows the established practice in DNA storage error correction literature \citep{press2020}, which tested error-correcting codes across DNA error rates ranging from $1\%$ to $15\%$ and demonstrated that such codes can achieve error-free recovery even at error rates up to $10\%$. Our grid spans this validated range to assess the robustness of pathway identification to the read-level error threshold.

	To ensure the resulting survival model has adequate statistical power while avoiding degenerate or saturated outcome definitions, we restricted the candidate threshold pairs to those satisfying three prespecified criteria: (1)~the baseline quality-failure rate must not exceed 15\%, ensuring that the outcome retains meaningful temporal variation beyond day~0; (2)~to avoid threshold combinations with extremely sparse event information, we required at least 100 observed storage-quality failures. This criterion is used as a pragmatic stability requirement rather than as a formal power guarantee; and (3)~the censoring rate must not exceed 80\%, to maintain a practically useful amount of event information for model estimation.

	From the 55 candidate pairs formed by the grid $\tau \in \{0.01, 0.03, 0.05, 0.07, 0.10\}$ and $\phi \in \{0.02, 0.025, 0.03, 0.05, 0.10, 0.15, 0.20, 0.25, 0.30, 0.40, 0.50\}$, six pairs satisfied all three criteria and were retained for further evaluation. For each admissible pair $(\tau,\phi)$, $n$~denotes the number of oligos retained after excluding baseline quality failures. Event counts and censoring rates are calculated within the retained analysis set.

		\begin{table}[H]
			\centering
			\caption{Admissible working points for the storage-quality failure thresholds in the Genscript pool.}
			\label{tab:tau-sensitivity}
			\small
			\begin{tabular}{ccccc}
				\hline
				$\tau$ & $\phi$ & Events & Censoring rate & $n$ \\
				\hline
				$0.03$ & $0.30$ & $1414$ & $77.16\%$ & $6190$ \\
				\hline
				$0.05$ & $0.15$ & $1341$ & $77.72\%$ & $6020$ \\
				\hline
				$0.10$ & $0.020$ & $4257$ & $31.68\%$ & $6231$ \\
				\hline
				$0.10$ & $0.025$ & $3633$ & $42.54\%$ & $6323$ \\
				\hline
				$0.10$ & $0.030$ & $3091$ & $51.68\%$ & $6397$ \\
				\hline
				$0.10$ & $0.050$ & $1465$ & $77.69\%$ & $6567$ \\
				\hline
			\end{tabular}
		\end{table}

\begin{figure}[H]
			\centering
			\includegraphics[width=1\textwidth]{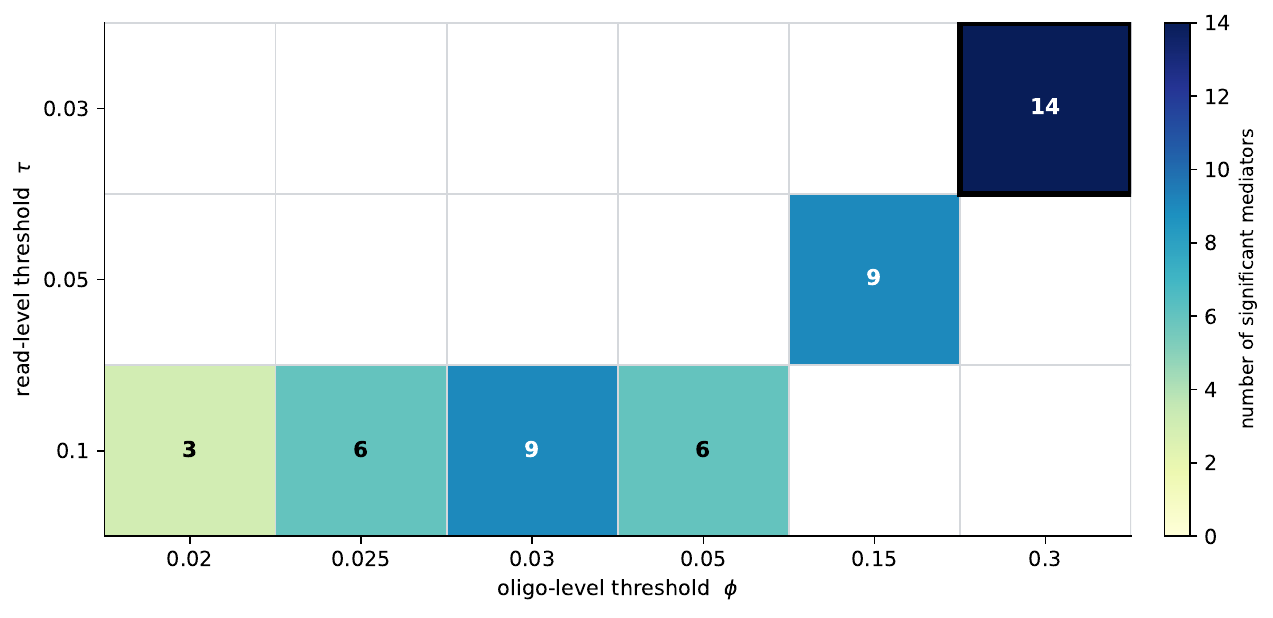}
			\caption{Sensitivity of pathway selection to the read-level threshold $\tau$ and oligo-level threshold $\phi$ in the Genscript data. Each cell reports the number of significant mediators; blank cells indicate threshold pairs that did not satisfy the admission criteria. The bold outline marks the primary analysis setting ($\tau=0.03$, $\phi=0.30$).}
			\label{fig:tau_phi}
		\end{figure}

\begin{figure}[htp]
			\centering
			\includegraphics[width=1\textwidth]{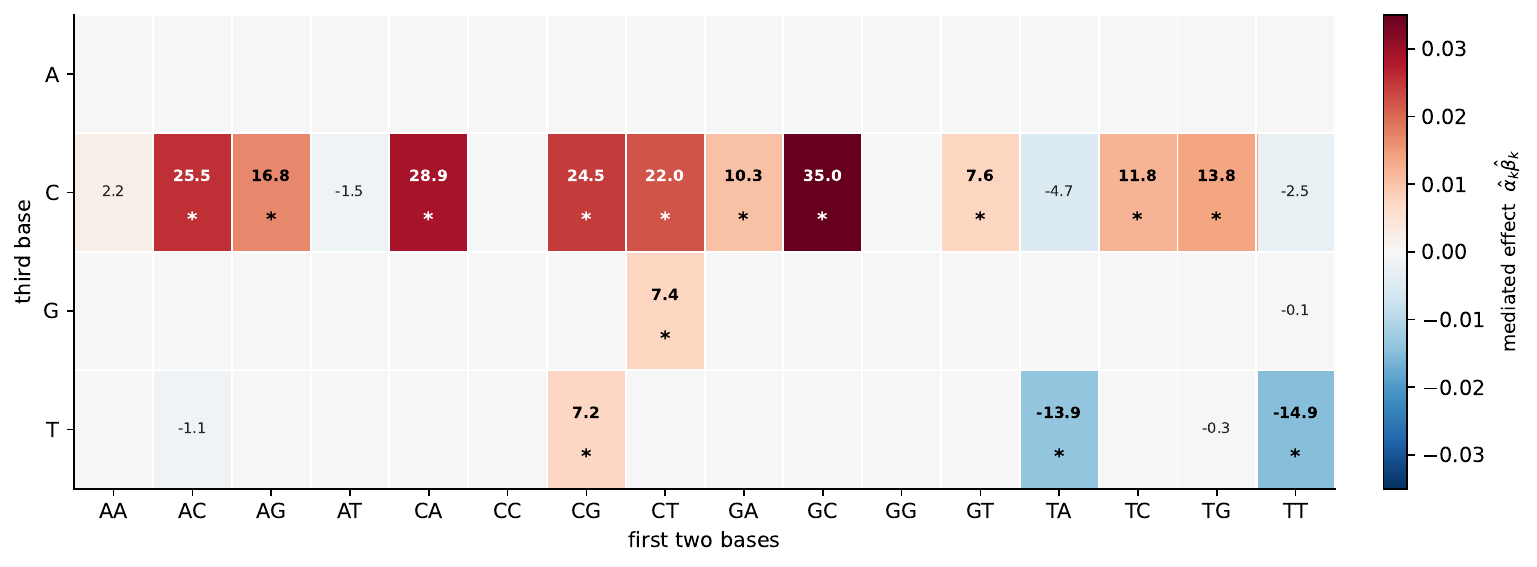}
			\caption{Context-specific mediated effects of single-base deletion pathways for the Genscript working point $(\tau=0.03,\phi=0.30)$. Rows and columns index the third and first two bases of each trinucleotide context, respectively. Cell values show $\hat{\alpha}_k\hat{\beta}_k \times 10^{3}$; color indicates the direction and magnitude of the estimated mediated effect.}
			\label{fig:deletion_heatmap}
		\end{figure}

		As summarized in Figure~\ref{fig:tau_phi}, we examined all $6$ admissible $(\tau, \phi)$ combinations. The qualitative finding that deletion-type mediators dominate the mediated pathway remained consistent. For example, at $\tau=0.03$ with $\phi=0.30$, all $14$ significant mediators were single-base deletions. The estimated direct log-hazard coefficient \(\widehat\gamma\) remained positive across all working points, indicating a consistently positive residual association between GC content and failure hazard after adjustment for the selected mediators. The mediated component was predominantly attributable to deletion-type pathways, although individual context-specific indirect effects could be positive or negative. Figure~\ref{fig:deletion_heatmap} further shows the context-specific mediated effects.

		At the working point $(\tau,\phi)=(0.03,0.30)$, the SCAD fit used $\lambda=0.04$ of which 14 passed the joint-significance test. In the heatmap below, empty cells therefore represent mediators whose fitted $\widehat{\beta}_k$ was shrunk to zero by SCAD rather than missing observations.

\begin{figure}[htp]
			\centering
			\includegraphics[width=0.8\textwidth]{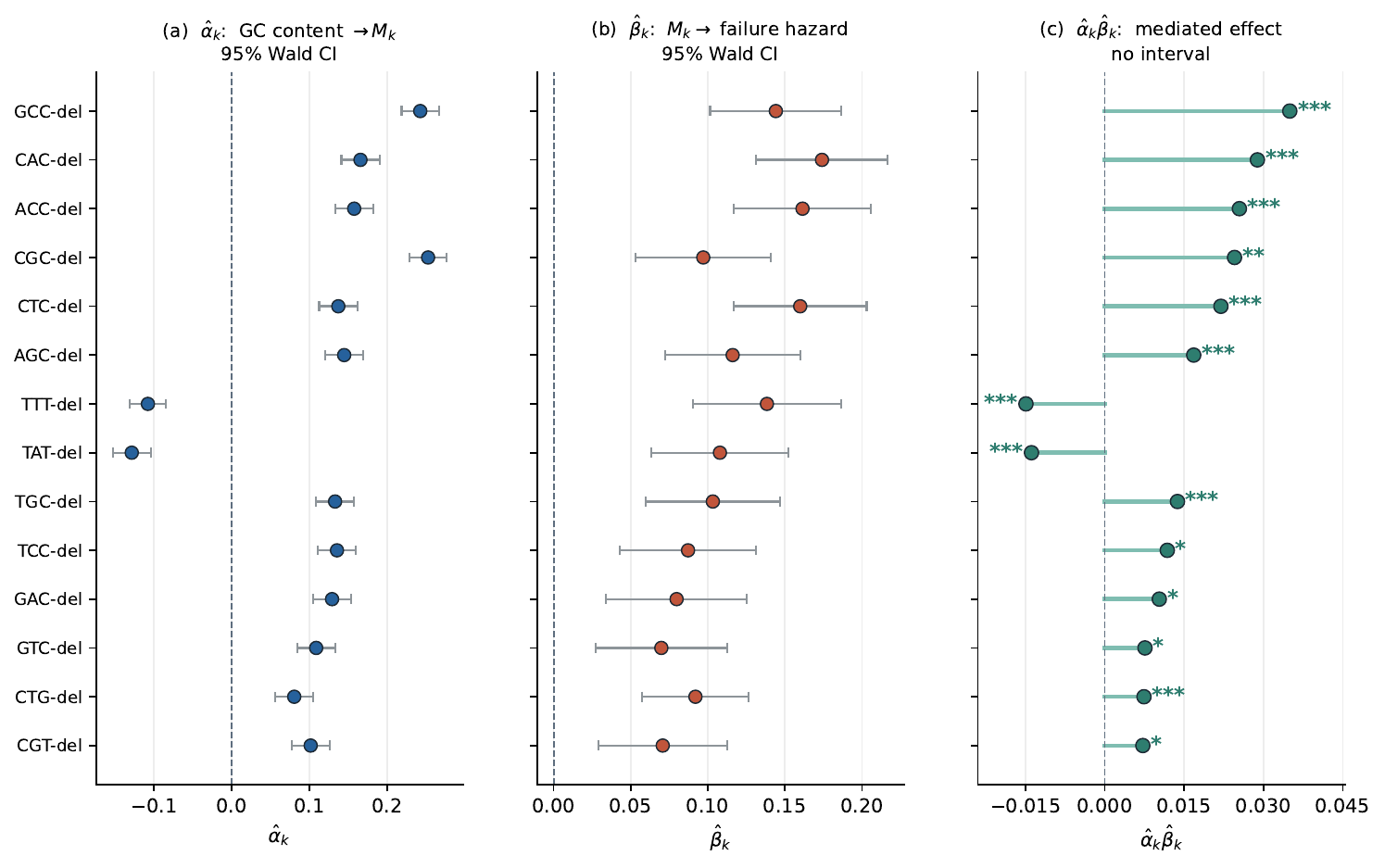}
			\caption{Pathway decomposition of the effect of GC content on storage failure. (a) and (b) show point estimates and 95\% Wald confidence intervals (CIs) for the exposure-to-mediator and mediator-to-outcome coefficients, respectively. (c) shows point estimates of the coefficient products without confidence intervals. Wald CIs in (a) and (b) are constructed as $\hat{\theta} \pm 1.96 \times \text{SE}(\hat{\theta})$ based on asymptotic normality. Mediator ordering is identical across panels (descending $|\hat{\alpha}_k \hat{\beta}_k|$). Significance levels: *** $p_{\text{joint}} < 10^{-4}$, ** $p_{\text{joint}} < 10^{-3}$, and * $p_{\text{joint}} < 0.05$.}
			\label{fig:decomp}
		\end{figure}

		Figure~\ref{fig:decomp} decomposes the same working point into its two constituent coefficient sets, reporting the exposure-to-mediator and mediator-to-outcome estimates separately alongside their products.

\end{document}